%% file: DHBdecays10-31-12.tex
\documentclass[aps]{revtex4}
\usepackage{graphicx,epsfig,rawfonts,pictex,multirow,longtable,booktabs}

\def\etal{{\it et al.}}

\def\hlf{{{1\over2}}}
\def\thlf{{{3\over2}}}
\def\fhlf{{{5\over2}}}
\def\shlf{{{7\over2}}}
\def\>{\rangle}
\def\<{\langle}

\def\rmb#1{{\bf #1}}

\def\beq{\begin{equation}}
\def\eeq{\end{equation}}

\def\beqy{\begin{eqnarray}}
\def\eeqy{\end{eqnarray}}

\begin{document}

\pagenumbering{arabic}

\title
{Heavy Diquark Symmetry Constraints for Strong Decays}
\author{B. Eakins and W. Roberts}
\affiliation{Department of Physics, Florida State University, Tallahassee, FL 32306}
\begin{abstract}
The Heavy Diquark Symmetry (HDS) of Doubly Heavy Baryons (DHBs) provides new insights into the spectroscopy of these hadrons.  We derive the consequences of this symmetry for the mass spectra and the decay widths of DHBs.  We compare these symmetry constraints to results from a nonrelativistic quark model for the mass spectra and results from the $^3P_0$ model for strong decays.  The quark model we implement was not constructed with these symmetries and contains interactions which explicitly break HDS.  Nevertheless these symmetries emerge.  We argue that the $^3P_0$ model and any other model for strong transitions which employs a spectator assumption explicitly respects HDS.  We also explore the possibility of treating the strange quark as a heavy quark and apply these ideas to $\Xi$, $\Xi_c$, and $\Xi_b$ baryons.  
\end{abstract}
\maketitle

\section{Introduction}
The spectroscopy of baryons with two heavy quarks (Doubly Heavy Baryons or DHBs) offers new insights into the heavy quark limit of QCD.  The remarkable feature of these particles is that they contain two distinct subsystems with very different properties.  There is a heavy-heavy subsystem made up of the two heavy quarks (a heavy diquark), and a heavy-light subsystem made up of the heavy diquark and the light degrees of freedom.  Interactions in the heavy diquark subsystem occur in the non-relativistic limit of QCD, and show many similarities to heavy quarkonium \cite{eakinsroberts,gershtein00}, while the heavy-light subsystem has a spectrum very similar to singly heavy mesons \cite{eakinsroberts,SavageSpectrum}.  Even more remarkable is the possibility that there is a symmetry between singly heavy mesons and the heavy-light subsystem of DHBs.  Savage and Wise first proposed a superflavor symmetry which relates heavy mesons to DHBs that contain a heavy diquark in the ground state \cite{SavageSpectrum}.  They argued that the Coulombic part of the confining potential dominates in the heavy quark limit making the excitation energy for the heavy diquark proportional to $M_Q \alpha^2_s(M_Q)$ in analogy to positronium.  Thus, in the heavy quark limit these excitations are much higher in energy than excitations of the light degrees of freedom.  However, it is clear from the spectra of charmonium and bottomonium that this is not the case for systems with $c$ and $b$ quarks and that the excitation energy of the heavy diquark should be approximately independent of the heavy quark mass for $\Xi_{cc}$, $\Xi_{bc}$, and $\Xi_{bb}$ baryons \cite{eakinsroberts}.  Recently, it has been suggested that the superflavor symmetry may be extended to include excited heavy diquarks \cite{eakinsroberts}.  This extension proposes that the interactions between the heavy diquark and the light degrees of freedom are independent of the excitation state, total angular momentum, and flavor of the heavy diquark.  The fingerprint of this symmetry should be apparent in the spectra and the strong decays of these states.  The goal of this work is to briefly review the consequences of this Heavy Diquark Symmetry (HDS) for DHB spectroscopy and examine the implications for strong decays.  

We are aware of  only one treatment of DHB strong decays in the literature.  Hu and Mehen \cite{Hu} use a Heavy Hadron Chiral Perturbation Theory (HH$\chi$PT) Lagrangian which incorporates the superflavor symmetry of Savage and Wise to calculate strong decay widths of the lowest lying $\Xi_{cc}$ and $\Xi_{bb}$ excited states.  They treat these lowest lying excited states as excitations of the heavy diquark (as we do).  These decays are suppressed at leading order, but $1/M_Q$ corrections are included in their treatment.  

In ref.~\cite{eakinsroberts} we discussed six aspects of the  heavy diquark symmetry.  In this note we explore the consequences of four of these for the strong decays of double heavy baryons. The first feature is that the decoupling of the total angular momentum of the heavy diquark leads to a spectrum that consists of degenerate multiplets similar to those of the heavy quark symmetry \cite{IsgurWise}. The consequence of this for strong decays is that the set of decays (by pion-emission, say) from the states of one multiplet to the states of another are related essentially by spin-counting coefficients.  The second feature is a heavy diquark selection rule that forbids transitions by light meson emission that alter the state of the heavy diquark.  We next examine the expectation that the strong decay amplitudes should be independent of the  heavy diquark's excitation state,  as well as its flavor. Finally, we explore the possibility of treating the strange quark as a heavy quark. The possible consequences of the superflavor symmetry are left for future work.  In order to accomplish these tasks, we use results from a constituent quark model (the Roberts-Pervin or RP model \cite{RobertsPervin}) for the spectra and results from the $^3P_0$ model for the transitions.  We also explore the features of these models which complement the heavy diquark symmetry and those features which are in tension with the symmetry.

\subsection{Heavy Diquarks as Meaningful Objects} \label{bloviations}

It is well known that for singly heavy hadrons the physics of the heavy quark (its spin and flavor) decouples from the physics of the light degrees of freedom \cite{IsgurWise}.  This occurs because interactions involving the heavy quark's spin occur through  its chromomagnetic moment ($\propto 1/M_Q$) which vanishes in the heavy quark limit.  The flavor symmetry arises because interactions which alter the heavy quark velocity are suppressed by $1/M_Q$, $M_Q \gg \Lambda_{QCD}$.  Thus, the heavy quark may be regarded as a stationary source of the color field.  However, if a hadron contains two heavy quarks, the heavy quarks may exchange a hard gluon which would alter their velocities.  This necessitates the inclusion of a heavy quark kinetic energy term in the effective Lagrangian which explicitly depends on the heavy quark mass, breaking the flavor symmetry at leading order \cite{Thacker91}.  On the other hand, if the two heavy quarks form a pointlike object (a heavy diquark), the dynamics of the heavy diquark decouple from the dynamics of the light degrees of freedom.  In this case a three-body problem (in a constituent quark model picture) `factorizes' into two, independent two-body problems \cite{eakinsroberts,kiselev02,gershtein00}. In this limit, the heavy diquark may be treated as a stationary source of the color field, and a heavy diquark flavor symmetry emerges.  The superflavor symmetry follows from this and the fact that the heavy diquark has the same color structure as a heavy antiquark.

If the size of the heavy diquark is much less than $1/\Lambda_{QCD}$, factorization is assured.  However, it may not be necessary for the diquark to be pointlike for this type of decoupling to emerge.  A simple example of this is a nonrelativistic system of three particles confined by harmonic oscillator potentials.  The Hamiltonian factorizes explicitly after a suitable choice of coordinates, provided that the masses of the heavy quarks are equal \cite{Castro}.  The RP model does not assume factorization nor is the mean separation between the two heavy quarks small.  Nevertheless, many of the consequences of factorization emerge in the DHB spectrum \cite{eakinsroberts}.  Remarkably, these consequences also appear to emerge for $\Xi$ baryons, if the strange quark is treated as heavy.  These results may be surprising because the mean separation between the two strange quarks is similar to the mean separation between the center of mass of the strange diquark and the light quark.  One can therefore conclude that, for this particular model, a pointlike diquark is not necessary for some approximate factorization to emerge.  The extent to which these systems factorize in QCD is a question requiring experimental investigation.

If factorization does indeed emerge for DHBs, then excitations of the light degrees of freedom should not mix significantly with excitations of the heavy diquark, implying that resonances may be labeled by the diquark quantum numbers and the quantum numbers for the light degrees of freedom.  Because heavy diquarks are not directly observable, the primary challenge with this approach is identifying the spectroscopy of heavy diquarks using DHB spectroscopy.  The first tool for accomplishing this task is a `supercolor symmetry' relating the spectra of heavy quarkonium to the spectra of heavy diquarks \cite{eakinsroberts}.  This analysis indicates the energy of an orbital excitation of a heavy diquark is about 230 MeV and nearly independent of flavor for doubly heavy hadrons containing $c$ and/or $b$ quarks.  One can also apply the superflavor symmetry to heavy meson spectroscopy to estimate that the energy of an orbital excitation of the light degrees of freedom is about 400 MeV for DHBs.  This leads to the na\"ive expectation that the lowest lying negative parity DHBs should consist primarily of an orbital excitation of the heavy diquark.  However, mass splittings alone do not provide a complete picture of the structure of the states. The pattern of strong decay amplitudes can prove to be a more useful tool for deciphering the structure of the states, particularly the amount of mixing between heavy diquark excitations and excitations of the light degrees of freedom.  In this work we illustrate that pion-emission decays of DHB resonances provide a reliable way of determining if a resonance contains an excited heavy diquark.

\section{Heavy Diquark Symmetry Constraints}\label{constraints}
\subsection{Total Angular Momentum Decoupling}
In this discussion we use the successful Heavy Quark Symmetry (HQS) \cite{IsgurWise} as a guide for exploring the symmetries which might emerge for doubly heavy hadrons.  For hadrons containing a single, infinitely massive heavy quark, the spin of the heavy quark decouples from the spin of the light degrees of freedom.  Thus, singly heavy hadrons exist in degenerate doublets with total angular momentum $J=J_l \pm s_Q$, where $s_Q=\frac{1}{2}$ is the spin of the heavy quark and $J_l$ is the total angular momentum of the light degrees of freedom.  For doubly heavy hadrons, the analogous expectation is that the total angular momentum of the heavy diquark decouples from the total angular momentum of the light degrees of freedom leading to degenerate multiplets with total angular momenta $J$ satisfying
\beq
|J_d-J_l| \le J \le J_d + J_l,
\eeq
\noindent where $J_d$ is the total angular momentum of the heavy diquark.  When applied to $J^{\pi_d}_d=1^+$, the symmetry proposed by Savage and Wise \cite{SavageSpectrum} is recovered. However, their formalism may be easily extended to diquarks with higher spin \cite{GeorgiFlav}.

\subsection{Heavy Diquark Spin-Counting}\label{hdsc}
The decoupling of the total angular momentum of the heavy diquark means that the strong decays of DHBs belonging to the same multiplet will be related by spin-counting arguments analogous to those of the Heavy Quark Symmetry (HQS) \cite{IsgurWise}.  For this work, we restrict the discussion to pion-emission.  The central idea behind these spin-counting arguments is that the decay process only involves the light degrees of freedom and the heavy diquark is a spectator.  This means that the total angular momentum of the light degrees of freedom is conserved in the process. For the decay $A\to B+C$, conservation of angular momentum at the hadronic level may be expressed as
\begin{equation}
\rmb{J}_l^\prime+\rmb{J}_d=\rmb{J}_b,\,\,\,
\rmb{J}_b+\rmb{J}_c^\prime=\rmb{J}_a,
\end{equation}
where $J_d$ is the angular momentum of the heavy diquark and $J_l^\prime$ is the total angular momentum of the light degrees of freedom in the daughter heavy baryon $B$. In addition, $\rmb{J}_c^\prime=\rmb{J}_c+{\bf \ell}$, with $J_c$ being the angular momentum of the light meson emitted in the decay, and $\ell$ is the orbital angular momentum between the light meson $C$ and the heavy daughter baryon $B$.  The amplitude for the decay described in this way may be written as ${\cal M}(J_a \rightarrow [J_b J_c^\prime]_{J_a})$.

Conservation of angular momentum at the level of the light degrees of freedom in the baryons may be expressed as
\begin{equation}
\rmb{J}_l^\prime+\rmb{J}_c^\prime=\rmb{J}_l,\,\,\,
\rmb{J}_l+\rmb{J}_d=\rmb{J}_a,
\end{equation}
where $J_l$ is the total angular momentum of the light degrees of freedom in the parent baryon $A$.  In this case, we write the amplitude as $A(J_d J_l \rightarrow J_d [J^\prime_{l} J_c^\prime]_{J_l})$. These two descriptions of the decay process are related by a Wigner $6-j$ symbol, and the amplitude for the process can be written as
\beq \label{HDS}
{\cal M}(J_a \rightarrow [J_b J_c^\prime]_{J_a})=(-1)^{J_d+J^\prime_l+J_c^\prime+J_a}\sqrt{(2J_b+1)(2J_l+1)}\left\{ \begin{array} {ccc} J_d & J^\prime_l   & J_b \\ 
                            J_c^\prime & J_a & J_l  \end{array} \right\} A^{J_c^\prime}_{J_l,J^\prime_l}.
\eeq
The HDS amplitude $A^{J_c^\prime}_{J_l,J^\prime_l}$ is independent of the total angular momentum, flavor, and excited state of the heavy diquark in the heavy quark limit.  The coefficients of $A^{J_c^\prime}_{J_l,J^\prime_l}$ are identical to those that occur in HQS if the total angular momentum of the heavy diquark $J_d$ is replaced with the spin of a heavy quark $S_Q=\hlf$.  From this expression it is possible to predict ratios of partial decay widths for states in an excited multiplet.  One consequence of the spin-counting relations is that total decay rates will be equal \cite{IsgurWise} for heavy mesons and DHBs.  This follows from the completeness relation for the $6-j$ symbol:
\beq
\sum_{J_b} (2J_b+1)(2J_l+1) \left(\left\{ \begin{array} {ccc} S_Q & J^\prime_l   & J_b \\ 
                            J_c^\prime & J_a & J_l  \end{array} \right\} \right)^2=1=\sum_{J_b} (2J_b+1)(2J_l+1) \left(\left\{ \begin{array} {ccc} J_d & J^\prime_l   & J_b \\ 
                            J_c^\prime & J_a & J_l  \end{array} \right\} \right)^2.
\eeq
This property implies that total decay widths of states belonging to the same multiplet will be equal, if  all channels (both real and virtual) are included and differences in phase space are taken into account.

\subsection{Heavy Diquark Selection Rule}
Because decays by light meson emission should occur completely within the light degrees of freedom in the baryon, the process cannot alter the state of the heavy diquark.  In strong decay models where the heavy diquark is treated as a spectator, such as the $^3P_0$ model, this selection rule is explicitly enforced (see section \ref{selectrule}).  If the spectator assumption is realized by QCD, then transitions between baryons in which the diquarks are (predominantly) in different quantum states can only occur through components of the wave function in which the diquarks are in the same quantum state.  Therefore, doubly heavy baryon resonances that do not decay by light meson emission to the ground-state multiplet must contain an excited heavy diquark.  This could provide useful information on the structure of doubly heavy baryons.

\subsection{Diquark Excitation and Flavor Symmetries}
It is expected  that the physics of the light degrees of freedom are independent of the excited state of the heavy diquark. As a result,  the energy required to excite the light degrees of freedom are expected to be independent of the heavy diquark's excitation state.  For instance, it has been argued in \cite{eakinsroberts} that the energy required to orbitally excite the light degrees of freedom from a $J^{\pi_l}_l=\hlf^+$ state to a $J^{\pi_l}_l=\hlf^-$ state is the same for a $J_d^{\pi_d}=1^+$ diquark as it is for a $J_d^{\pi_d}=1^-$ diquark, about 350 MeV in the RP model \cite{RobertsPervin}. As seen by the light component of the baryon,  an excited heavy diquark appears simply as a heavier version of the ground state diquark, with different total angular momentum and parity. Thus, the strong decay widths of states containing diquarks with $J_d^{\pi_d}=1^+$ should be equal to the corresponding strong decay widths of states containing diquarks with $J_d^{\pi_d}=1^-$, after phase space differences are taken into account.  These relations test the validity of factorization (see section \ref{bloviations}) for systems which contain an excited heavy diquark.

It is also expected that the properties of the light degrees of freedom of a DHB are independent of the flavor of the heavy diquark, implying that a host of strong decay amplitudes among the $\Xi_{cc}$, $\Xi_{bc}$, and the $\Xi_{bb}$ will be related in a simple way.  It has been pointed out that the flavor symmetry of HQS is broken in heavy-heavy systems because one cannot neglect the kinetic energy of the heavy quarks \textit{a priori} in an effective field theory approach \cite{Thacker91}.  However, there is some indication from models that, to an approximation, the heavy-light subsystem factorizes from the heavy-heavy subsystem allowing one to treat each subsystem separately \cite{eakinsroberts,kiselev02,gershtein00}.  In this way it is possible to think of the heavy diquark as a stationary source of anti-color, so that a diquark flavor symmetry emerges in the heavy-light subsystem, despite the violation of the heavy quark flavor symmetry in the heavy-heavy subsystem.  The dependence of the decay rates on the heavy diquark's mass will enter primarily in the phase space for the decay.  This phase space factor is divided off when the strong decay amplitudes of baryons containing different flavors of diquarks are compared.

\section{Quark Model}
\subsection{Hamiltonian}
The work presented here utilizes the wave functions from a non-relativistic quark model which Roberts and Pervin \cite{RobertsPervin} used to calculate masses of heavy baryons.  The advantage of this model is that it is among only a handful of articles \cite{RobertsPervin, Ebert:2002ig, gershtein00, Majethiya} in which orbital excitations both in the heavy diquark and in the light degrees of freedom are allowed.  Among these models, it is the only one which solves the three-body problem without using a quark-diquark approximation.  The Hamiltonian used includes a confining potential,
\beq
V^{ij}_{\rm conf}= \sum_{i<j=1}^3\left({br_{ij}\over 2}-
{2\alpha_{\rm Coul}\over3r_{ij}}\right);
\eeq
\noindent with $r_{ij}=\vert\rmb{r}_i-\rmb{r}_j \vert$;  a hyperfine interaction,
\beq\label{hyp}
H^{ij}_{\rm hyp} =\sum_{i<j=1}^3\left[{2\alpha_{\rm con}\over 3 m_i 
m_j}{8\pi\over 3} \rmb{S}_i\cdot\rmb{S}_j\delta^3(\rmb{r}_{ij})
+ {2\alpha_{\rm tens}\over 3m_i m_j}{1\over {r}^3_{ij}}\left(
{3\rmb{S}_i\cdot\rmb {r}_{ij} \rmb{S}_j\cdot\rmb {r}_{ij}\over {r}^2_{ij}} 
-\rmb{S}_i\cdot\rmb{S}_j\right) \right];
\eeq
\noindent and an ad-hoc spin orbit interaction,
\beq \label{SO}
V_{\rm SO} = \frac{\alpha_{\rm SO}}{\rho^2+\lambda^2} \frac{\rmb{L}\cdot\rmb{S}}{(m_1+m_2+m_3)^2},
\eeq
\noindent in which $L$ is the total orbital angular momentum of the baryon, and $S$ is the total spin of the three quarks.  The spin orbit term explicitly depends on the Jacobi coordinates, 
\beq
{\bf \rho}= \frac{1}{\sqrt2}({\bf r}_1 - {\bf r}_2)
\eeq
and 
\beq
{\bf \lambda} =\sqrt{\frac{2}{3}}\left(\frac{m_1{\bf r}_1 + m_2{\bf r}_2}{m_1+m_2} - 
{\bf r}_3\right).
\eeq
The parameters used in the model are listed in Table \ref{parameter1}.

\begin{center}
\begin{table}[h]
\caption{Parameters of the RP model \cite{RobertsPervin}.  $m_{\sigma}$ is the mass of the $u$ and $d$ quarks.\label{parameter1}}
\vspace{5mm}
\begin{tabular}{ccccccccc}
 \hline \toprule
 $m_\sigma$ & $m_s$  & $m_c$  & $m_b$  & $b$& $\alpha_{\rm Coul}$ 
 &$\alpha_{\rm con}$&$\alpha_{\rm SO}$ &$\alpha_{\rm tens}$   \\
 (GeV)&(GeV)&(GeV)&(GeV)&(GeV$^2$)& && (GeV) & \\ \hline
0.2848& 0.5553&  1.8182 & 5.2019 &  0.1540 &$\approx 0.0$&1.0844 &0.9321& -0.2230 \\ \hline
\end{tabular}
\end{table}
\end{center}

The radial part of the wave functions are solved for by expanding in a basis of harmonic oscillator wave functions, $\psi_{nl}$, in each Jacobi coordinate;  
\beq
\Psi_{n_\rho l_\rho, n_\lambda l_\lambda} (\rho, \lambda)= \psi_{n_\rho l_\rho}(\alpha_\rho \rho) \psi_{n_\lambda l_\lambda}(\alpha_\lambda \lambda).
\eeq
\noindent The size parameters, $\alpha_\rho$ and $\alpha_\lambda$, are determined using a variational calculation, and the basis is expanded up to the second harmonic oscillator band, $2(n_\rho + n_\lambda)+l_\rho+l_\lambda \le 2$.  

For the purposes of this note, we also consider two modifications of this model. The first modification is a spinless version of the model.  Because the heavy diquark is a composite object, the decoupling of its total angular momentum does not necessarily arise in the limit of infinitely massive heavy quarks (see section \ref{pairmix}).  Removing spin dependent interactions enables us to determine how well the diquark total angular momentum decoupling emerges in this limit.  The second modification appends the spinless model with a spin orbit term of the form
\beq \label{microSO}
V_{\rm \mu SO}=a \emph{ {\bf l}}_\lambda \cdot {\bf s}_l,
\eeq
where $a$ is a constant chosen arbitrarily, $\rmb{l}_\lambda$ is the relative orbital angular momentum between the light quark and the heavy diquark, and $\rmb{s}_l=\hlf$ is the spin of the light quark.  An interaction of this type is not suppressed in the heavy quark limit, and it may have important consequences for the mixing of states.  The spin orbit interaction in the original model, Eq.~(\ref{SO}) is not sensitive to the microscopic dynamics of DHB systems. Inclusion of the additional term above, Eq.~(\ref{microSO}), would remedy this.  Therefore, the second modification of the model removes all spin interactions which depend on the mass of a heavy quark and includes this surviving microscopic spin orbit interaction.  As such, it should be a more accurate representation of the heavy quark limit than the spinless model.  We also remark that this interaction does not include the appropriate spatial dependence, and the value of the constant has not been fit to experimental data.  However, we are primarily concerned with how a term of this type will mix quark model states, so this simplified implementation of a full spin orbit interaction is sufficient for our purposes.

\subsection{Flavor Multiplets}\label{flavormultiplets}

Assuming that only the $b$ and $c$ quarks can be treated as heavy, states containing two heavy quarks can be placed in SU(2) multiplets. These multiplets are a triplet
\begin{equation}
 \Xi_{cc}=ccq,\,\,\,\,\,\Xi_{bb}=bbq,\,\,\,\,\Xi_{bc}=\frac{1}{\sqrt{2}}\left(cb+bc\right)q,
\end{equation}
and a singlet
\begin{equation}
\Xi^\prime_{bc}=\frac{1}{\sqrt{2}}\left(cb-bc\right)q.
\end{equation}
Of course, this is only meant to be a classification scheme, but it has implications for the angular momentum structure of the states, since the spin-space-flavor wave function must be symmetric in `identical' quarks. For instance, if there is no excitation of the heavy diquark, the diquarks in the flavor triplet must have spin 1, while those in the flavor singlet must have spin 0.  

\subsection{Angular Momentum Multiplets}\label{Multiplets}
In the quark model presented above, states are defined in what we call an $L-S$ basis.  Symbolically, the angular momenta in the states are coupled in the scheme $\{[l_\rho l_\lambda]_L [s_d s_l]_S\}_J$, where $l_\rho$ is the orbital angular momentum between the two heavy quarks, $l_\lambda$ is the relative angular momentum between the light quark and the heavy diquark, $s_d$ is the spin of the heavy diquark, and $s_l=\frac{1}{2}$ is the spin of the light quark.  For the symmetry scheme discussed in section \ref{constraints}, states are identified by the spin and parity of the diquark, $J_{d}^{\pi_d}$, the spin and parity of the light degrees of freedom, $J_{l}^{\pi_l}$, and the total angular momentum and parity of the baryon, $J^P$.  In order to compare these states with quark model states it is necessary to transform the states from the $L-S$ basis to the $J_d-J_l$ basis, in which the angular momenta are coupled symbolically as $\{[l_\rho s_d]_{J_d} [l_\lambda s_l]_{J_l}\}_J$.  These two bases are related by a $9-j$ symbol,
\beq\label{HQSSstates}
 \sqrt{(2L+1)(2S+1)(2J_d+1)(2J_l+1)}
\left\{ \begin{array} {ccc} l_\rho & l_\lambda   & L \\ 
                            s_d & \frac{1}{2} & S \\
                            J_d & J_l   & J \end{array} \right\}.
\eeq
\noindent Using this, we can assign states obtained in the quark model to angular momentum supermultiplets by examining their wave function components.  

\subsection{Pair-wise Mixing}\label{pairmix}
HDS is broken in this model even when spin-dependent interactions are absent.  This arises because any pair-wise interaction that is a function of $|\rmb{r}_{ij}|$ will be diagonal in the $L-S$ basis. This means that such an interaction will necessarily mix states in the $J_d-J_l$ basis.  In the absence of spin-dependent interactions, the mixing coefficients are independent of the mass of the heavy quarks and the strength of the interaction. In general, this mixing also leads to a mass splitting between states that occupy the same symmetry multiplet, and thereby lifts the degeneracy of states within the multiplet. This is discussed in more detail in \cite{eakinsroberts}.

\section{$^{3}P_{0}$ Model}
\subsection{Transition Operator}
For this work, we use the $^{3}P_{0}$ model to calculate decay amplitudes \cite{Roberts-Brac,LeY73,LeY74}.  In this model, strong decays proceed through quark-antiquark pair creation.  The created antiquark combines with one of the quarks from the parent baryon to form the daughter meson, and the created quark combines with the other two quarks (the heavy diquark in this case) to form the daughter baryon. The {\it ansatz} for the transition operator is 
\begin{eqnarray}
T =-3g\sum_{i,j} \int d{\bf p_{i}} d{\bf p_{j}} \delta({\bf p_{i}}+{\bf p_{j}}) C_{ij} F_{ij}
\sum_{m} \< 1,m;1,-m|0,0\> \chi^{m}_{ij} {\cal Y}^{-m}_{1}({\bf p_{i}}-{\bf p_{j}})b^{\dag}_{i}({\bf p_{i}}) d^{\dag}_{j}({\bf p_{j}})  ,
\end{eqnarray}
where $g$ is a dimensionless constant, and $b^{\dag}_{i}({\bf p_{i}})$  and $d^{\dag}_{j}({\bf p_{j}})$ are the creation operators for a quark and antiquark with momenta ${\bf p_{i}}$ and ${\bf p_{j}}$, respectively. This pair is characterized by a color singlet wave function $C_{ij}$, a flavor wave function $F_{ij}$, and a spin wave function
$\chi _{ij}^m$ with total spin $S$ = 1 and $z$-projection $m$.   The phenomenological constant  $g$ is chosen to have the value $1.4$ in order to reproduce the decay width of the $\Xi(1530)$ (9.1$\pm$0.5 MeV).  For the daughter pion, we use a single component ground state harmonic oscillator wave function with a size parameter $\beta=0.623$ GeV.

\subsection{Selection Rule and Spin-Counting} \label{selectrule}
For decays to a light daughter meson, the two heavy quarks of the parent DHB are spectators in the process, so that the transition operator is proportional to an overlap of the heavy diquark ($\rho$ coordinate) part of the DHB wave functions.  The orthogonality of these wave functions implies that the transition does not alter the total spin, relative orbital angular momentum, radial quantum number or total angular momentum of the heavy diquark.  This implies that the $^{3}P_{0}$ model explicitly forbids transitions between DHBs with different diquark quantum numbers, provided that the wave functions are HDS eigenstates.  This should be a feature of any nonrelativistic decay model which includes a spectator assumption.

There is also a selection rule for the light degrees of freedom.  To demonstrate this, we write the transition operator in a  general form
\begin{eqnarray}
T \propto \{{\cal O}_{J_d'} \otimes  {\cal O}_{J_l'}\}_{J'},
\end{eqnarray}
${\cal O}_{J_d'}$ is a spherical tensor operator of rank ${J_d'}$ acting only on the heavy diquark degrees of freedom, and ${\cal O}_{J_l'}$ is a spherical tensor operator of rank ${J_l'}$ acting only on the light degrees of freedom (in principle this operator should be a sum over all possible spherical tensors, but this complication is irrelevant to this discussion).  The only way in which angular momentum conservation is guaranteed is if $J'=0$, which means that $J_d'=J_l'$.  As we have explained above, the spectator assumption requires that $J_d'=0$, thus $J_l'=0$, and the total angular momentum of the light degrees of freedom is conserved.  These two conservation rules imply that the transition matrix elements are proportional to the expected spin-counting factor (Eq.~(\ref{HDS})).  We should mention that these arguments also imply that transition models with a spectator assumption also produce the appropriate HQS spin-counting factor for singly heavy hadron decays.  We derive this feature of the $^{3}P_{0}$ model for both DHBs and heavy mesons in the appendix.  Thus, we have argued that the $^{3}P_{0}$ model and any other model with a spectator assumption explicitly respects HDS and HQS constraints.

\subsection{Phase Space} \label{PhaseSpace}
Using Fermi's golden rule, the decay width for a parent baryon $A$ in its rest frame, decaying to daughter baryon $B$, and a daughter meson $C$, is
\beq
\Gamma_{A\rightarrow BC}(l)=2\pi \int dk_0 k_0^2 \delta[m_a - E_b(k_0) - E_c(k_0)] |M_{A \rightarrow BC} (k_0,l)|^2,
\eeq
with
\beq
E_b(k_0)=\sqrt{k_0^2+m_b^2}, \, E_c(k_0)=\sqrt{k_0^2+m_c^2}.
\eeq
Here $l$ is the relative orbital angular momentum of the daughter hadrons, $k_0$ is the magnitude of the relative three momentum between the two daughter hadrons (calculated in the rest frame of the parent), $M_{A \rightarrow BC }(k_0,l)$ is the transition amplitude, and $m_a$, $m_b$, $m_c$ are the masses of particles $A$, $B$, and $C$, respectively. The final expression for the decay width is
\beq
\Gamma_{A\rightarrow BC}(l)=2\pi \frac{E_b(k_0)E_c(k_0)k_0}{m_a} |M_{A \rightarrow BC} (k_0,l)|^2,
\eeq
When comparing decay rates to symmetry constraints it is useful to divide off the phase space factors in order to remove effects due to the broken degeneracy of multiplets. However, this does not remove all of the effects of broken degeneracies as the amplitude is a function of $k_0$ which depends on the masses of the hadrons involved.  In any case, we divide off a phase space factor of 
\beq
\frac{2\pi E_b(k_0)E_c(k_0)k_0^{2l+1}}{m_a}
\eeq
from the calculated decay widths to obtain the square of what we henceforth refer to as the ``decay amplitude''.

\section{Results}

Tables \ref{tripwidths} to \ref{MSOsingwidths} show the results we obtain using the $^3P_0$ operator to calculate the pion-emission amplitudes/widths of the DHBs. Tables \ref{Chitripwidths} to \ref{XiMSOsingwidths} contain analogous results obtained for systems with strange diquarks by treating the strange quark as a heavy quark. Various features of these results are discussed in some detail in the next few subsections.

\subsection{Angular Momentum Multiplets and Decay Widths} \label{DHBwidths}

The pion-emission decay widths of the flavor-triplet $\Xi_{cc}$, $\Xi_{bc}$, and $\Xi_{bb}$ states shown in Table \ref{tripwidths}. All decay widths shown in that table are for decays to the ground state $n_d^{2s_d+1}(l_\rho)_{J_d}\otimes n_\lambda(l_\lambda)_{J_l}=1^{3}S_1\otimes 1S_\frac{1}{2}$. The analogous decays of the flavor-singlet $\Xi_{bc}^\prime$ to the flavor-singlet ground-state  $n_d^{2s_d+1}(l_\rho)_{J_d}\otimes n_\lambda(l_\lambda)_{J_l}=1^{1}S_0\otimes 1S_\frac{1}{2}$ are shown in Table \ref{singwidths}.

In Table \ref{tripwidths}, the first column shows the quantum numbers of the parent baryon on the first line, and the total angular momentum and parity of the states comprising the nearly-degenerate multiplet on the second line. Column two shows the masses of the corresponding  states for the $\Xi_{cc}$. Column three shows a matrix of decay widths, with the first row being the decay widths to the ground-state daughter baryon with $J^P=\hlf^+$ and the second row containing the decay widths to the ground-state daughter baryon with $J^P=\thlf^+$. The elements in each row of the matrix are the pion-emission decay widths of each member of the parent multiplet to the particular member of the ground-state daughter multiplet. Columns four and five present the masses and decay widths of the $\Xi_{bc}$, respectively, in the same format. Columns six and seven present the same information, in the same format, for the $\Xi_{bb}$. For the decays of the flavor-singlet states shown in Table \ref{singwidths}, the numbers in columns two and three are also presented in the same format. The $J^P$ multiplets shown in these two tables are identified and discussed in \cite{eakinsroberts}.
 
Many of the widths shown in these two tables are quite small (some are less than 1 keV), because these states are primarily composed of excited heavy diquarks: the spectator assumption of the $^3P_0$ model forbids these transitions.  For such states, the electromagnetic decays will therefore be quite important.  For flavor-triplet states, the largest of these suppressed decay modes occurs for the $\hlf^+$ member of the $1^{1}P_1\otimes 1P_\frac{3}{2}$ multiplet decaying to the $\thlf^+$ member of the ground state doublet.  This width is 6.90 MeV for the $\Xi_{cc}$, largely due to various small mixings in the wave functions. This corresponding  width for the $\Xi_{bb}$ drops to 0.512 MeV.  For the $\Xi^\prime_{bc}$ the largest of the suppressed decay modes is the decay of the $\hlf^+$ member of the $1^{3}P_2\otimes 1P_\frac{3}{2}$ multiplet to the ground state singlet (1.40 MeV). 

\begin{center}
\begin{table}[h]
\caption{Decay widths for pion-emission from the parent multiplet listed to the ground state $n_d^{2s_d+1}(l_\rho)_{J_d}\otimes  n_\lambda(l_\lambda)_{J_l}=1^{3}S_1\otimes 1S_\frac{1}{2}$ doublet for flavor-triplet states.  The first column shows the quantum numbers of the parent baryon on the first line, and the total angular momentum and parity of the states comprising the nearly-degenerate multiplet on the second line. Column two shows the masses of the corresponding states for the $\Xi_{cc}$. Column three shows a matrix of decay widths, with the first row being the decay widths to the ground-state daughter baryon with $J^P=\hlf^+$ and the second row containing the decay widths to the ground-state daughter baryon with $J^P=\thlf^+$. The elements in each row of the matrix are the pion-emission decay widths of each member of the parent multiplet to the particular member of the ground-state daughter multiplet. Columns four and five present the masses and decay widths of the $\Xi_{bc}$, respectively, in the same format. Columns six and seven present the same information, in the same format, for the $\Xi_{bb}$.  These decay widths are calculated with the $^3P_0$ model using quark model wave functions taken from the RP model \cite{RobertsPervin}.
\label{tripwidths}}
\vspace{2mm}
\resizebox{7.07in}{!} {
\begin{tabular}{c c c c c c c}
\hline \toprule
Parent Multiplet, $J_a^{P_a}$   & \multicolumn{6}{c }{Masses (GeV) and Widths (MeV)} \\ 
$n_d^{2s_d+1}(l_\rho)_{J_d}\otimes  n_\lambda(l_\lambda)_{J_l}$ & \multicolumn{2}{c}{\large{$\Xi_{cc}$}} & \multicolumn{2}{c}{\large{$\Xi_{bc}$}}& \multicolumn{2}{c}{\large{$\Xi_{bb}$}} \\ \cmidrule{2-3} \cmidrule{4-5} \cmidrule{6-7}
& Mass (GeV) & Width (MeV)  &Mass (GeV) & Width (MeV)& Mass (GeV) & Width (MeV)\\\hline

$\begin{array}{c}1^{1}P_1 \otimes  1S_\frac{1}{2}\\  (\hlf^-,\thlf^-)\end{array}$   & (3.911, 3.917)   
& $\left( \begin{array} {c} 0.658 \:\: 0.000 \\ 0.000 \:\: 0.297 \end{array} \right)$  & (7.212, 7.214)
& $\left( \begin{array} {c} 0.210 \:\: 0.000 \\ 0.000 \:\: 0.069 \end{array} \right)$  & (10.470, 10.470)
& $\left( \begin{array} {c} 0.015 \:\: 0.000 \\ 0.000 \:\: 0.000 \end{array} \right)$ \vspace{4mm} \\ 

$\begin{array}{c}2^{3}S_1 \otimes  1S_\frac{1}{2} \\ (\hlf^+,\thlf^+)\end{array}$        & (4.030, 4.078)
& $\left( \begin{array} {c} 0.021 \:\: 0.016 \\ 0.238 \:\: 0.005 \end{array} \right)$  & (7.321, 7.353)
& $\left( \begin{array} {c} 0.011 \:\: 0.004 \\ 0.122 \:\: 0.004 \end{array} \right)$  & (10.551, 10.574)
& $\left( \begin{array} {c} 0.001 \:\: 0.000 \\ 0.005 \:\: 0.001 \end{array} \right)$ \vspace{4mm} \\  

$\begin{array}{c}1^{3}D_1\otimes  1S_\frac{1}{2} \\   (\hlf^+,\thlf^+)\end{array}$    & (4.098, 4.045\footnotemark[1])       
& $\left( \begin{array} {c} 0.018 \:\: 0.020 \\ 0.024 \:\: 0.000 \end{array} \right)$  & (7.372, 7.337\footnotemark[1])
& $\left( \begin{array} {c} 0.010 \:\: 0.008 \\ 0.008 \:\: 0.000 \end{array} \right)$  & (10.589, 10.564\footnotemark[1])
& $\left( \begin{array} {c} 0.002 \:\: 0.001 \\ 0.001 \:\: 0.000 \end{array} \right)$ \vspace{4mm} \\ 

$\begin{array}{c}1^{3}D_2\otimes  1S_\frac{1}{2}  \\  (\thlf^+,\fhlf^+)\end{array}$  & (4.094\footnotemark[1], 4.092)
& $\left( \begin{array} {c} 0.074 \:\: 0.000 \\ 0.017 \:\: 0.016 \end{array} \right)$  & (7.369\footnotemark[1], 7.368)
& $\left( \begin{array} {c} 0.033 \:\: 0.000 \\ 0.005 \:\: 0.008 \end{array} \right)$  & (10.587\footnotemark[1], 10.588)
& $\left( \begin{array} {c} 0.007 \:\: 0.000 \\ 0.001 \:\: 0.002 \end{array} \right)$ \vspace{4mm} \\ 

$\begin{array}{c}1^{3}D_3\otimes  1S_\frac{1}{2}\\  (\fhlf^+,\shlf^+)\end{array}$ & (4.048, 4.095)
& $\left( \begin{array} {c} 0.000 \:\: 0.000 \\ 0.021 \:\: 0.000 \end{array} \right)$  & (7.340, 7.370)
& $\left( \begin{array} {c} 0.000 \:\: 0.000 \\ 0.010 \:\: 0.000 \end{array} \right)$  & (10.568, 10.588)
& $\left( \begin{array} {c} 0.000 \:\: 0.000 \\ 0.000 \:\: 0.000 \end{array} \right)$ \vspace{4mm} \\ 

$\begin{array}{c}1^{3}S_1\otimes  1P_\frac{1}{2} \\  (\hlf^-,\thlf^-)\end{array}$   & (4.081, 4.077)
& $\left( \begin{array} {c} 396.0 \:\: 0.972 \\ 0.095 \:\: 277.6 \end{array} \right)$  & (7.397, 7.392)
& $\left( \begin{array} {c} 409.1 \:\: 1.606 \\ 0.071 \:\: 396.8 \end{array} \right)$  & (10.694, 10.691)
& $\left( \begin{array} {c} 423.3 \:\: 0.519 \\ 0.058 \:\: 328.9 \end{array} \right)$ \vspace{4mm} \\ 

$\begin{array}{c}1^{3}S_1\otimes  1P_\frac{3}{2} \\  (\hlf^-,\thlf^-,\fhlf^-)\end{array}$ & (4.073, 4.079, 4.089)
& $\left( \begin{array} {c} 2.050 \:\: 8.653 \:\: 29.15 \\ 14.62 \:\: 94.58 \:\: 9.149 \end{array} \right)$  & (7.390, 7.394, 7.399)
& $\left( \begin{array} {c} 1.646 \:\: 6.438 \:\: 23.10 \\ 17.69 \:\: 27.22 \:\: 9.587 \end{array} \right)$  & (10.691, 10.692, 10.695)
& $\left( \begin{array} {c} 0.926 \:\: 6.725 \:\: 19.98 \\ 22.21 \:\: 92.17 \:\: 10.99 \end{array} \right)$ \vspace{4mm} \\ 

$\begin{array}{c}1^{1}P_1\otimes  1P_\frac{1}{2} \\  (\hlf^+,\thlf^+)\end{array}$  & (4.257\footnotemark[2], 4.253)
& $\left( \begin{array} {c} 0.180 \:\: 0.111 \\ 0.002 \:\: 0.018 \end{array} \right)$  & (7.555\footnotemark[2], 7.549)
& $\left( \begin{array} {c} 0.105 \:\: 0.053 \\ 0.001 \:\: 0.014 \end{array} \right)$  & (10.817\footnotemark[2], 10.812)
& $\left( \begin{array} {c} 0.034 \:\: 0.007 \\ 0.000 \:\: 0.001 \end{array} \right)$ \vspace{4mm} \\ 

$\begin{array}{c}1^{1}P_1\otimes  1P_\frac{3}{2}  \\  (\hlf^+,\thlf^+,\fhlf^+)\end{array}$& (4.230\footnotemark[2], 4.261, 4.259)
& $\left( \begin{array} {c} 0.979 \:\: 0.043 \:\: 0.048 \\ 6.898 \:\: 0.002 \:\: 0.017 \end{array} \right)$  & (7.534\footnotemark[2], 7.557, 7.549)
& $\left( \begin{array} {c} 0.360 \:\: 0.026 \:\: 0.018 \\ 2.060 \:\: 0.001 \:\: 0.016 \end{array} \right)$  & (10.802\footnotemark[2], 10.818, 10.811)
& $\left( \begin{array} {c} 0.115 \:\: 0.008 \:\: 0.004 \\ 0.512 \:\: 0.000 \:\: 0.000 \end{array} \right)$ \vspace{4mm} \\ 

$\begin{array}{c}1^{3}S_1\otimes  2S_\frac{1}{2} \\ (\hlf^+,\thlf^+)\end{array}$ & (4.311, 4.368)    
& $\left( \begin{array} {c} 1.856 \:\: 0.118 \\ 38.41 \:\: 2.809 \end{array} \right)$  & (7.634, 7.676)
& $\left( \begin{array} {c} 1.568 \:\: 0.016 \\ 29.42 \:\: 3.099 \end{array} \right)$  & (10.940, 10.972)
& $\left( \begin{array} {c} 1.557 \:\: 0.457 \\ 22.32 \:\: 2.962 \end{array} \right)$ \vspace{4mm} \\ 

$\begin{array}{c}1^{3}S_1\otimes  1D_\frac{3}{2} \\ (\hlf^+,\thlf^+,\fhlf^+)\end{array}$   & (4.394, 4.394, 4.387)
& $\left( \begin{array} {c} 126.4 \:\: 82.01 \:\: 3.321 \\ 16.38 \:\: 90.51 \:\: 79.26 \end{array} \right)$  & (7.709, 7.708, 7.701
& $\left( \begin{array} {c} 129.2 \:\: 83.81 \:\: 0.139 \\ 16.98 \:\: 70.31 \:\: 137.4 \end{array} \right)$  & (11.011,  11.011, 11.002)
& $\left( \begin{array} {c} 135.9 \:\: 82.27 \:\: 6.052 \\ 17.35 \:\: 117.2 \:\: 48.74 \end{array} \right)$ \vspace{4mm} \\ 

$\begin{array}{c}1^{3}S_1\otimes  1D_\frac{5}{2}  \\  (\thlf^+,\fhlf^+,\shlf^+)\end{array}$  & (4.391, 4.388, 4.393)
& $\left( \begin{array} {c} 1.384 \:\: 28.39 \:\: 78.03 \\ 71.65 \:\: 144.6 \:\: 39.65 \end{array} \right)$  & (7.706, 7.702, 7.708)
& $\left( \begin{array} {c} 0.002 \:\: 32.49 \:\: 81.15 
                         \\ 108.6 \:\: 97.03 \:\: 46.26 \end{array} \right)$  & (11.011,  11.002, 11.011)
& $\left( \begin{array} {c} 4.784 \:\: 26.91 \:\: 82.90 
                         \\ 77.42 \:\: 192.6 \:\: 52.51 \end{array} \right)$ \vspace{4mm} \\ \hline
\toprule \footnotetext[1]{$^{\text{b}}$These states mix significantly.}
\end{tabular}}
\end{table}
\end{center}
\begin{center}
\begin{table}[h]
\caption{Decay widths for pion-emission from the parent multiplet listed to the ground state $n_d^{2s_d+1}(l_\rho)_{J_d}\otimes  n_\lambda(l_\lambda)_{J_l}=1^{1}S_0\otimes  1S_\frac{1}{2}$ singlet for flavor-singlet states. The first column shows the quantum numbers of the parent baryon on the first line, and the total angular momentum and parity of the states comprising the nearly-degenerate multiplet on the second line. Column two shows the masses of the corresponding  states for the $\Xi_{bc}^\prime$. Column three shows the decay widths of each member of the parent multiplet.
\label{singwidths}}
\vspace{2mm}
\renewcommand{\arraystretch}{1.0}
\resizebox{4in}{!}{\begin{tabular}{c c c }
\hline \toprule
Parent Multiplet              &\multicolumn{2}c{ $\Xi^\prime_{bc}$} \\ 
$n_d^{2s_d+1}(l_\rho)_{J_d}\otimes  n_\lambda(l_\lambda)_{J_l}$ &Mass (GeV) & Width (MeV)\\ \hline

$\begin{array}{c}1^{3}P_0\otimes 1S_\frac{1}{2}\\ (\hlf^-)\end{array}$  &  (7.230\footnotemark[1])    & (0.068)  \vspace{1mm} \\ 

$\begin{array}{c}1^{3}P_1\otimes 1S_\frac{1}{2}\\(\hlf^-,\thlf^-)\end{array}$   &  (7.199\footnotemark[1], 7.228)  & (0.173  0.000)   \vspace{1mm}\\ 

$\begin{array}{c}1^{3}P_2\otimes 1S_\frac{1}{2}\\(\thlf^-,\fhlf^-)\end{array}$  &(7.201, 7.265)   & (0.000  0.000)   \vspace{1mm}\\ 

$\begin{array}{c}2^{1}S_0\otimes 1S_\frac{1}{2}\\(\hlf^+)\end{array}$ &   (7.333)  & (0.015)   \vspace{1mm}\\ 

$\begin{array}{c}1^{1}D_2\otimes 1S_\frac{1}{2}\\(\thlf^+,\fhlf^+)\end{array}$  & (7.361, 7.367)   & (0.020  0.000)  \vspace{1mm} \\

$\begin{array}{c}1^{1}S_0\otimes 1P_\frac{1}{2}\\(\hlf^-)\end{array}$ &  (7.388)   & (399.0)  \vspace{1mm}\\ 

$\begin{array}{c}1^{1}S_0\otimes 1P_\frac{3}{2}\\(\thlf^-)\end{array}$   &   (7.390)  & (28.06) \vspace{1mm} \\ 

$\begin{array}{c}1^{3}P_0\otimes 1P_\frac{1}{2}\\(\hlf^+)\end{array}$ &   (7.548\footnotemark[2])  & (0.015)  \vspace{1mm}\\ 

$\begin{array}{c}1^{3}P_0\otimes 1P_\frac{3}{2}\\(\thlf^+)\end{array}$ &   (7.546\footnotemark[3])  & (0.059) \vspace{1mm} \\ 

$\begin{array}{c}1^{3}P_1\otimes 1P_\frac{1}{2}\\(\hlf^+,\thlf^+)\end{array}$ &  (7.552\footnotemark[2], 7.556\footnotemark[3]) & (0.003  0.034) \vspace{1mm} \\ 

$\begin{array}{c}1^{3}P_1\otimes 1P_\frac{3}{2}\\(\hlf^+,\thlf^+,\fhlf^+)\end{array}$   & (7.555\footnotemark[2], 7.548\footnotemark[3], 7.553\footnotemark[4])  & (0.033  0.000  0.001)   \vspace{1mm} \\ 

$\begin{array}{c}1^{3}P_2\otimes 1P_\frac{1}{2}\\(\thlf^+,\fhlf^+)\end{array}$ &   (7.563\footnotemark[3], 7.548\footnotemark[4])  & (0.016  0.000)  \vspace{1mm} \\ 

$\begin{array}{c}1^{3}P_2\otimes 1P_\frac{3}{2}\\(\hlf^+,\thlf^+,\fhlf^+,\shlf^+)\end{array}$   & (7.519\footnotemark[2], 7.562\footnotemark[3], 7.559\footnotemark[4], 7.557) & (1.399  0.002  0.000  0.009)  \vspace{1mm} \\ 

$\begin{array}{c}1^{1}S_0\otimes 2S_\frac{1}{2}\\(\hlf^+)\end{array}$ &   (7.645)  & (11.10)  \vspace{1mm} \\ 

$\begin{array}{c}1^{1}S_0\otimes 1D_\frac{3}{2}\\(\thlf^+)\end{array}$ & (7.709)   & (133.6)   \vspace{1mm}\\ 

$\begin{array}{c}1^{1}S_0\otimes 1D_\frac{5}{2}\\(\fhlf^+)\end{array}$  &  (7.689)  & (113.3)  \vspace{.5mm} \\\hline

\toprule
\footnotetext[1]{$^{\text{bcd}}$These states mix significantly.}
\end{tabular}}
\end{table}
\end{center}

There are also decay modes in Table \ref{tripwidths} whose suppression arise not from the spectator overlap, but from spin-counting arguments.  In Table \ref{tripwidths} for instance, the decay width of the $\hlf^-$ member of the $1^{3}S_1\otimes 1P_\frac{1}{2}$ multiplet to the ground state $\thlf^+$ is only 0.095 MeV for the $\Xi_{cc}$.  This particular decay can only occur in a D-wave because of the conservation of total angular momentum and parity, but the $6-j$ symbol in Eq.~(\ref{HDS}) vanishes for this partial wave.  The suppression is easily understood by noting that the total angular momentum of the light degrees of freedom is conserved in the decay process.  The total angular momentum of the light degrees of freedom in the daughter baryon with $J^\prime_l=\hlf$ must therefore couple to relative orbital angular momentum $l$ to yield the total angular momentum of the light degrees of freedom for the parent baryon $J_l=\hlf$.  This condition cannot be met with $l=2$, so this decay only occurs through the small mixing between the $J^P=\hlf^-$ states of the  $1^{3}S_1\otimes 1P_\frac{1}{2}$ and $1^{3}S_1\otimes 1P_\frac{3}{2}$ multiplets.

The decay widths for the allowed decays of the $1^{3}S_1\otimes 1P_\frac{1}{2}$ multiplet (Table~\ref{tripwidths}) and the $1^{1}S_0\otimes 1P_\frac{1}{2}$ multiplet (Table~\ref{singwidths}) are quite large.  For these states, the S-wave decay mode is allowed by HDS.  In the RP model these are the lowest lying states with excitations of the light degrees of freedom.  If these states are indeed this broad, then it might present a problem for identifying the members of these multiplets experimentally. However, it is known that the $^3P_0$ model leads to unusually large $S$-wave widths, and this may simply be another instance of that feature. 

We close this subsection by noting that the hyperfine interaction mixes the $\Xi_{bc}$ and $\Xi^\prime_{bc}$ states. It has been shown that this mixing significantly affects the semileptonic decays of the ground state $\Xi_{bc}$ abd $\Xi_{bc}^\prime$ \cite{Hypmix}. Similar effects can be expected among the strong decay amplitudes, but we defer a more detailed discussion of this to possible future work.  

\subsubsection{HDS Spin-Counting} \label{ResultsSpinCount}

Table \ref{SpinCount1+} shows  ratios of decay amplitudes for states in which the heavy diquark is in its ground state. These are calculated in three variants of the RP model: the full model, the spinless model, and the $\mu$SO model, which is the spinless RP model appended with an additional $\rmb{l}_\lambda \cdot \rmb{s}_l$ interaction.  The appropriate spin-counting factors have been divided out so that each of these ratios should be unity.  In the full model most of these ratios are in reasonable agreement with this expectation, even though the ratios of amplitudes are expected to be sensitive to mixing. Not surprisingly, the ratios obtained in the full RP model get closer to spin-counting expectations as the mass of the heavy diquark increases. The worst disagreement in the full model occurs when the light quark is radially excited in the $1 ^3S_1\otimes 2S_\hlf$ multiplet. This is probably because for radially excited states there is a zero in the wave function of the initial state, leading to a zero in the amplitude for the decay. The position of this zero is very sensitive to the size parameters of the wave functions as well as the relative momentum of the final state particles, $k_0$.  This can lead to significant departures from the HDS predictions. The full model also mixes this multiplet with the $1 ^3S_1\otimes 1S_\hlf$ ground state multiplet, but this mixing is larger for the $\hlf^+$ member of the $1 ^3S_1\otimes 2S_\hlf$ multiplet than it is for the $\thlf^+$ member.  This problem with the $1 ^3S_1\otimes 2S_\hlf$ multiplet's spin-counting ratios vanishes in the spinless RP model as well as in the $\mu$SO model where the states in each multiplet are completely degenerate (so $k_0$ is the same for all decays, and the position of the node in the amplitude does not depend on the particular decay).
\begin{center}
\begin{table}[h]
\caption{ratios of decay amplitudes for states in which the heavy diquark is in its ground state. These are calculated in three variants of the RP model: the full model, the spinless model, and the $\mu$SO model, which is the spinless RP model appended with an additional $\rmb{l}_\lambda \cdot \rmb{s}_l$ interaction.  The appropriate spin-counting factors have been divided out so that each of these ratios should be unity.
\label{SpinCount1+}}
\vspace{5mm}
\renewcommand{\arraystretch}{1.9}
\begin{tabular}{cc|ccc|ccc|ccc}

\hline \toprule
Multiplet & Decay Ratios & \multicolumn{3}{c|}{Full Model} & \multicolumn{3}{c|}{Spinless}& \multicolumn{3}{c}{$\mu$SO} \vspace{-3pt} \\ 
$n_d^{2s_d+1}(l_\rho)_{J_d}\otimes n_\lambda(l_\lambda)_{J_l}$ &${\cal M}(J_a,J_b,l)$ & $\Xi_{cc}$&$\Xi_{bc}$& $\Xi_{bb}$& $\Xi_{cc}$& $\Xi_{bc}$& $\Xi_{bb}$ & $\Xi_{cc}$& $\Xi_{bc}$& $\Xi_{bb}$ 
 \\  \hline

\multirow{1}{*}{$1^{3}S_1\otimes 1P_\hlf$} &\large{$\frac{{\cal M}(\thlf,\thlf,0)}{{\cal M}(\hlf,\hlf,0)}$}  &
                               1.110 & 1.019 & 1.034 & 0.791 & 0.790 & 0.791 & 1.004 & 0.996 & 1.002 \\[+4pt] \hline

\multirow{4}{*}{$1 ^3S_1\otimes 1P_\thlf$} &\large{$\frac{1}{\sqrt{5}} \frac{{\cal M}(\thlf,\hlf,2)}{{\cal M}(\hlf,\thlf,2)}$}  &
                               0.827 & 0.811 & 0.907 & 0.791 & 0.791 & 0.791 & 1.000 & 0.999 & 1.000 \\[+4pt] 
                                 &\large{$\frac{2}{\sqrt{5}} \frac{{\cal M}(\thlf,\thlf,2)}{{\cal M}(\hlf,\thlf,2)}$} &
                               0.932 & 0.884 & 0.958 & 0.791 & 0.791 & 0.791 & 1.000 & 1.000 & 1.000 \\[+4pt] 
                                 &\large{$\sqrt{\frac{8}{15}} \frac{{\cal M}(\fhlf,\hlf,2)}{{\cal M}(\hlf,\thlf,2)}$} &   
                               0.860 & 0.900 & 0.938 & 1.061 & 1.061 & 1.061 & 1.000 & 0.999 & 1.000  \\[+4pt] 
                                 &\large{$\sqrt{\frac{7}{15}} \frac{{\cal M}(\fhlf,\thlf,2)}{{\cal M}(\hlf,\thlf,2)}$} &
                               0.976 & 0.983 & 0.991 & 1.061 & 1.061 & 1.061 & 1.000 & 0.999 & 1.000   \\[+4pt]  \hline

\multirow{3}{*}{$1 ^3S_1\otimes 2S_\hlf$} &\large{$\frac{1}{2\sqrt{2}} \frac{{\cal M}(\hlf,\thlf,1)}{{\cal M}(\hlf,\hlf,1)}$}  &
                               2.030 & 1.804 & 1.480 & 1.000 & 1.000 & 1.000 & 0.996 & 1.002 & 0.998 \\[+4pt] 
                                 &\large{$\frac{1}{2} \frac{{\cal M}(\thlf,\hlf,1)}{{\cal M}(\hlf,\hlf,1)}$} &
                               0.106 & 0.046 & 0.246 & 1.004 & 0.996 & 1.007 & 1.041 & 0.954 & 1.026 \\[+4pt] 
                                 &\large{$\frac{1}{\sqrt{5}} \frac{{\cal M}(\thlf,\thlf,1)}{{\cal M}(\hlf,\hlf,1)}$}&   
                               0.573 & 0.644 & 0.613 & 1.004 & 0.996 & 1.007 & 1.036 & 0.957 & 1.024  \\[+4pt]   \hline

\multirow{4}{*}{$1 ^3S_1\otimes 1D_\thlf$} &\large{$2\sqrt{2} \frac{{\cal M}(\hlf,\thlf,1)}{{\cal M}(\hlf,\hlf,1)}$}  &
                               1.234 & 1.184 & 1.105 & 1.001 & 1.000 & 1.001 & 1.005 & 0.995 & 1.003 \\[+4pt] 
                                 &\large{$2\sqrt{\frac{2}{5}} \frac{{\cal M}(\thlf,\hlf,1)}{{\cal M}(\hlf,\hlf,1)}$} &
                               1.020 & 1.020 & 0.983 & 0.895 & 0.894 & 0.895 & 1.004 & 0.995 & 1.003 \\[+4pt] 
                                 &\large{$\sqrt{2} \frac{{\cal M}(\thlf,\thlf,1)}{{\cal M}(\hlf,\hlf,1)}$} &   
                               1.259 & 1.203 & 1.084 & 0.895 & 0.894 & 0.895 & 1.009 & 0.990 & 1.005  \\[+4pt] 
                                 &\large{$\frac{2 \sqrt{2}}{3} \frac{{\cal M}(\fhlf,\thlf,1)}{{\cal M}(\hlf,\hlf,1)}$} &
                               1.212 & 1.208 & 1.004 & 0.721 & 0.720 & 0.726 & 0.994 & 0.975 & 0.989   \\[+4pt]  \hline

\multirow{4}{*}{$1 ^3S_1\otimes 1D_\fhlf$} &\large{$\sqrt{\frac{21}{5}} \frac{{\cal M}(\fhlf,\hlf,3)}{{\cal M}(\thlf,\thlf,3)}$}  &
                               0.771 & 0.864 & 0.864 & 0.837 & 0.843 & 0.852 & 1.027 & 1.032 & 1.047 \\[+4pt] 
                                 &\large{$\frac{\sqrt{21}}{4} \frac{{\cal M}(\fhlf,\thlf,3)}{{\cal M}(\thlf,\thlf,3)}$} &
                               0.965 & 1.016 & 0.958 & 0.837 & 0.843 & 0.852 & 1.026 & 1.033 & 1.046 \\[+4pt] 
                                 &\large{$\frac{\sqrt{7}}{2} \frac{{\cal M}(\shlf,\hlf,3)}{{\cal M}(\thlf,\thlf,3)}$} &   
                               0.805 & 0.847 & 0.930 & 1.118 & 1.118 & 1.118 & 1.001 & 0.999 & 1.000  \\[+4pt] 
                                 &\large{$\sqrt{\frac{7}{3}} \frac{{\cal M}(\fhlf,\thlf,2)}{{\cal M}(\hlf,\thlf,2)}$} &
                               1.000 & 0.990 & 1.024 & 1.118 & 1.118 & 1.118 & 1.000 & 1.000 & 1.000   \\[+4pt]\hline

 \toprule
\end{tabular}
\end{table}
\end{center}

The results for spin-counting ratios in the spinless RP model show a peculiar feature.  While many of the ratios are in better agreement with HDS than those for the full model, several of them get worse when spin-dependent interactions are turned off. One instance of this can be seen in the comparison of the results from the full model and the spinless model in the second and third rows of Table \ref{SpinCount1+}.  The heavy diquark spin symmetry appears to be broken in the model even though spin interactions have been removed.  The origin of this lies in the choice of basis and the mixing due to pairwise forces (see section \ref{pairmix}).  The spinless RP model was calculated using the L--S basis, and all of the states used in Table \ref{SpinCount1+} are diagonal in this basis.  This implies that the energy eigenstates are linear combinations of HDS eigenstates (the $J_d$--$J_l$ basis).  In principle the pairwise nature of the confining potential used in the RP model imposes this mixing of HDS eigenstates. However, the energy splittings associated with this mixing vanish unless the states contain both an orbitally excited heavy diquark {\it and} orbitally excited light degrees of freedom.  There is therefore some freedom in the choice of basis for these particular states.  None of the decays in Table \ref{SpinCount1+} contain excited heavy diquarks.  Therefore, the disagreement with HDS for these particular ratios is an artifact of any model which uses an L--S basis.  This can be seen by examining the results for the $\mu$SO model in Table \ref{SpinCount1+}.  Adding a small interaction to the spinless RP model of the form $\rmb{l}_\lambda \cdot \rmb{s}_l$ forces these states to be eigenstates in the $J_d$--$J_l$ basis and produces results in quite good agreement with the expectations of HDS.  An examination of the wave function components in the $\mu$SO model reveals that this modification of the spinless RP model removes nearly all of the mixing of the HDS states in Table \ref{SpinCount1+}.

Tables \ref{NoPStripwidths} and \ref{NoPSsingwidths} show the pion-emission decay amplitudes obtained using the full wave functions of the RP model, for a number of states, including some that contain orbitally excited diquarks. Tables \ref{MSOtripwidths} and \ref{MSOsingwidths} show the analogous numbers for the $\mu$SO model. In these tables,  the spin-counting factors have been divided out so that all of the amplitudes associated with a particular transition between HDS multiplets should be equal.  In the full RP model (Table \ref{NoPStripwidths}), spin-counting relations for transitions from the $1^{1}P_1\otimes 1P_\frac{1}{2}$ multiplet to the $1^{1}P_1\otimes 1S_\frac{1}{2}$ multiplet work reasonably well,  while the relations for transitions from the $1^{1}P_1\otimes 1P_\frac{3}{2}$ multiplet to the $1^{1}P_1\otimes 1S_\frac{1}{2}$ multiplet are only slightly worse. This results despite the mixing that occurs between states in these two multiplets. For flavor-singlet states in the full model (Table \ref{NoPSsingwidths}), these relations perform significantly more poorly, particularly for decays of the $1^{3}P_2\otimes 1P_\frac{3}{2}$ quadruplet to the $1^{3}P_2\otimes 1S_\frac{1}{2}$ doublet.  This is due to the significant amount of mixing that these states experience.  Mixing is also much more complicated for the flavor-singlet states than for the flavor-triplet states. In the flavor-triplet case most of the mixing that occurs is predominantly two-level mixing, but for the flavor-singlet case as many as five different multiplets may be involved in the mixing.
\begin{center}
\begin{table}[h]
\caption{HDS decay amplitudes of flavor-triplet states by pion-emission from the parent multiplet listed to the daughter multiplet listed.  The amplitudes are arranged in matrix form with the parent baryon's position in the multiplet corresponding to the column of the matrix and the daughter baryon's position in the multiplet corresponding to the row of the matrix.  These decay amplitudes are extracted from the $^3P_0$ model using quark model wave functions taken from the RP model \cite{RobertsPervin}.  Phase space has been divided off as well as the appropriate spin-counting factors.  Thus, all of the amplitudes, labeled by $A^l_{J_l, J^\prime_l}$, in each section of the table should be equal according to lowest order HDS predictions.  The entries in each set of parentheses are equal because of spin-counting arguments, the entries in different columns are equal because of the diquark flavor symmetry, and the entries involving different excitation states of the heavy diquark are equal because of the excitation symmetry.  The decay modes omitted are forbidden by spin-counting arguments.
\label{NoPStripwidths}}
\vspace{2mm}
\resizebox{7.07in}{!} {
\begin{tabular}{ccccccc}

\hline \toprule
\multirow{2}{*}{Parent Multiplet} & \multirow{2}{*}{$J_a^{P_a}$} & \multirow{2}{*}{Daughter Multiplet} & \multirow{2}{*}{$\qquad J_b^{P_b} \qquad$} & \multicolumn{3}{c}{HDS Amplitude $A^l_{J_l, J^\prime_l}$ (GeV$^{-l-\hlf}$)} \\ 
 & && & \large{$\Xi_{cc}$} & \large{$\Xi_{bc}$} & \large{$\Xi_{bb}$} \\ \hline
 & & & & \multicolumn{3}{c}{\large{$A^0_{\hlf, \hlf}$}} \vspace{1mm} \\ \cline{5-7} 

$1^{3}S_1\otimes 1P_\frac{1}{2}$ & $(\hlf^-,\thlf^-)$           
& $1^{3}S_1\otimes 1S_\frac{1}{2}$ & $\left( \begin{array} {c} \hlf^+ \\ \thlf^+ \end{array} \right)$
& $\left( \begin{array} {c} 0.708 \;\;  \quad-\quad  \\  \quad-\quad  \;\; 0.786 \end{array} \right)$  
& $\left( \begin{array} {c} 0.724 \;\;  \quad-\quad  \\  \quad-\quad  \;\; 0.738 \end{array} \right)$  
& $\left( \begin{array} {c} 0.748 \;\;  \quad-\quad  \\  \quad-\quad  \;\; 0.774 \end{array} \right)$ \vspace{2mm} \\  

$1^{1}P_1\otimes 1P_\frac{1}{2}$              & $(\hlf^+,\thlf^+)$           
& $1^{1}P_1\otimes 1S_\frac{1}{2}$ & $\left( \begin{array} {c} \hlf^- \\ \thlf^- \end{array} \right)$ 
& $\left( \begin{array} {c} 0.607 \;\;  \quad-\quad  \\  \quad-\quad  \;\; 0.710 \end{array} \right)$  
& $\left( \begin{array} {c} 0.605 \;\;  \quad-\quad  \\  \quad-\quad  \;\; 0.702 \end{array} \right)$  
& $\left( \begin{array} {c} 0.600 \;\;  \quad-\quad  \\  \quad-\quad  \;\; 0.689 \end{array} \right)$ \vspace{2mm} \\ 

 & & & & \multicolumn{3}{c}{\large{$A^2_{\thlf, \hlf}$}} \vspace{1mm} \\ \cline{5-7} 

$1^{3}S_1\otimes 1P_\frac{3}{2}$              & $(\hlf^-,\thlf^-,\fhlf^-)$           
& $1^{3}S_1\otimes 1S_\frac{1}{2}$ & $\left( \begin{array} {c} \hlf^+ \\ \thlf^+ \end{array} \right)$ 
& $\left( \begin{array} {c} \quad-\quad   1.826 \;\; 1.898 \\ 2.208 \;\; 2.057 \;\; 2.156 \end{array} \right)$  
& $\left( \begin{array} {c} \quad-\quad   1.724 \;\; 1.913 \\ 2.126 \;\; 1.879 \;\; 2.090 \end{array} \right)$  
& $\left( \begin{array} {c} \quad-\quad   1.857 \;\; 1.920 \\ 2.047 \;\; 1.960 \;\; 2.029 \end{array} \right)$ \vspace{2mm} \\ 

$1^{1}P_1\otimes 1P_\frac{3}{2}$              & $(\hlf^+,\thlf^+,\fhlf^+)$           
& $1^{1}P_1\otimes 1S_\frac{1}{2}$ & $\left( \begin{array} {c} \hlf^- \\ \thlf^- \end{array} \right)$ 
& $\left( \begin{array} {c} \quad-\quad   1.892 \;\; 2.125 \\ 1.771 \;\; 1.913 \;\; 2.144 \end{array} \right)$  
& $\left( \begin{array} {c} \quad-\quad   1.828 \;\; 2.087 \\ 1.702 \;\; 1.835 \;\; 2.093 \end{array} \right)$  
& $\left( \begin{array} {c} \quad-\quad   1.765 \;\; 2.033 \\ 1.626 \;\; 1.769 \;\; 2.037 \end{array} \right)$ \vspace{2mm} \\ 

 & & & & \multicolumn{3}{c}{\large{$A^1_{\hlf, \hlf}$}} \vspace{1mm} \\ \cline{5-7} 

$1^{3}S_1\otimes 2S_\frac{1}{2}$              & $(\hlf^+,\thlf^+)$           
& $1^{3}S_1\otimes 1S_\frac{1}{2}$ & $\left( \begin{array} {c} \hlf^+ \\ \thlf^+ \end{array} \right)$
& $\left( \begin{array} {c} 0.167 \;\;  0.018  \\  0.339  \;\; 0.095 \end{array} \right)$  
& $\left( \begin{array} {c} 0.145 \;\;  0.006  \\  0.261  \;\; 0.093 \end{array} \right)$  
& $\left( \begin{array} {c} 0.140 \;\;  0.034  \\  0.207  \;\; 0.086 \end{array} \right)$ \vspace{2mm} \\ 

 & & & & \multicolumn{3}{c}{\large{$A^1_{\thlf, \hlf}$}}  \vspace{1mm} \\ \cline{5-7}

$1^{3}S_1\otimes 1D_\frac{3}{2}$              & $(\hlf^+,\thlf^+,\fhlf^+)$           
& $1^{3}S_1\otimes 1S_\frac{1}{2}$ & $\left( \begin{array} {c} \hlf^+ \\ \thlf^+ \end{array} \right)$
& $\left( \begin{array} {c} 0.387 \;\; 0.395, \quad-\quad \\ 0.482 \;\; 0.488 \;\; 0.469 \end{array} \right)$  
& $\left( \begin{array} {c} 0.372 \;\; 0.379, \quad-\quad \\ 0.440 \;\; 0.447 \;\; 0.449 \end{array} \right)$  
& $\left( \begin{array} {c} 0.372 \;\; 0.366, \quad-\quad \\ 0.411 \;\; 0.403 \;\; 0.374 \end{array} \right)$ \vspace{2mm} \\ 

 & & & & \multicolumn{3}{c}{\large{$ A^3_{\fhlf, \hlf}$}}  \vspace{1mm} \\ \cline{5-7} 

$1^{3}S_1\otimes 1D_\frac{5}{2}$              & $(\thlf^+,\fhlf^+,\shlf^+)$           
& $1^{3}S_1\otimes 1S_\frac{1}{2}$ & $\left( \begin{array} {c} \hlf^+ \\ \thlf^+ \end{array} \right)$
& $\left( \begin{array} {c} \quad-\quad   0.876 \;\; 0.915 \\ 1.136 \;\; 1.096 \;\; 1.136 \end{array} \right)$  
& $\left( \begin{array} {c} \quad-\quad   0.887 \;\; 0.870 \\ 1.027 \;\; 1.044 \;\; 1.017 \end{array} \right)$  
& $\left( \begin{array} {c} \quad-\quad   0.786 \;\; 0.846 \\ 0.910 \;\; 0.872 \;\; 0.932 \end{array} \right)$ \vspace{2mm} \\ \hline

\toprule
\end{tabular}
}
\end{table}
\end{center}
\begin{center}
\begin{table}[h]
\caption{HDS decay amplitudes of flavor-singlet states by pion-emission from the parent multiplet listed to the daughter multiplet listed.  The amplitudes are arranged in matrix form with the parent baryon's position in the multiplet corresponding to the column of the matrix and the daughter baryon's position in the multiplet corresponding to the row of the matrix.  These decay amplitudes are extracted from the $^3P_0$ model using quark model wave functions taken from the RP model \cite{RobertsPervin}.  Phase space has been divided off as well as the appropriate spin-counting factors.  Thus, all of the amplitudes, labeled by $A^l_{J_l, J^\prime_l}$, in each section of the table should be equal according to lowest order HDS predictions.  The entries in each set of parentheses are equal because of spin-counting arguments, and the entries involving different excitation states of the heavy diquark are equal because of the excitation symmetry.  These entries should also equal the corresponding amplitudes for flavor-triplet states.  The decay modes omitted are forbidden by spin-counting arguments.
\label{NoPSsingwidths}}
\vspace{2mm}

\begin{tabular}{c c c c c}
\hline \toprule
\multirow{2}{*}{Parent Multiplet} & \multirow{2}{*}{$J_a^{P_a}$} & \multirow{2}{*}{Daughter Multiplet} & \multirow{2}{*}{$\qquad J_b^{P_b} \qquad$} & HDS Amplitude $A^l_{J_l, J^\prime_l}$ (GeV$^{-l-\hlf}$) \\ 
 & && &  \large{$\Xi^\prime_{bc}$}  \\ \hline
 & & & & \large{$A^0_{\hlf, \hlf}$} \vspace{1mm} \\ \cline{5-5} 

$1^{1}S_0\otimes 1P_\frac{1}{2}$              & $(\hlf^-)$           
& $1^{1}S_0\otimes 1S_\frac{1}{2}$ & $(\hlf^+)$ & (0.781) \vspace{1mm} \\ 

$1^{3}P_0\otimes 1P_\frac{1}{2}$              & $(\hlf^+)$           
& $1^{3}P_0\otimes 1S_\frac{1}{2}$ & $(\hlf^-)$ & (0.405) \vspace{1mm} \\ 

$1^{3}P_1\otimes 1P_\frac{1}{2}$              & $(\hlf^+,\thlf^+)$           
& $1^{3}P_1\otimes 1S_\frac{1}{2}$ & $\left( \begin{array} {c} \hlf^- \\ \thlf^- \end{array} \right)$
& $\left( \begin{array} {c} 0.322 \;\;  \quad-\quad  \\  \quad-\quad  \;\; 0.427 \end{array} \right)$ \vspace{2mm} \\ 

$1^{3}P_2\otimes 1P_\frac{1}{2}$              & $(\thlf^+,\fhlf^+)$           
& $1^{3}P_2\otimes 1S_\frac{1}{2}$ & $\left( \begin{array} {c} \thlf^- \\ \fhlf^- \end{array} \right)$
& $\left( \begin{array} {c} 0.451 \;\;  \quad-\quad  \\  \quad-\quad  \;\; 0.749 \end{array} \right)$ \vspace{2mm} \\ 

 & & & & \large{$A^2_{\thlf, \hlf}$}  \vspace{1mm} \\ \cline{5-5}

$1^{1}S_0\otimes 1P_\frac{3}{2}$              & $(\thlf^-)$           
& $1^{1}S_0\otimes 1S_\frac{1}{2}$ & $(\hlf^+)$ & (2.044) \vspace{1mm} \\ 

$1^{3}P_0\otimes 1P_\frac{3}{2}$              & $(\thlf^+)$           
& $1^{3}P_0\otimes 1S_\frac{1}{2}$ & $(\hlf^-)$ & (1.326) \vspace{1mm} \\ 

$1^{3}P_1\otimes 1P_\frac{3}{2}$              & $(\hlf^+,\thlf^+,\fhlf^+)$           
& $1^{3}P_1\otimes 1S_\frac{1}{2}$ & $\left( \begin{array} {c} \hlf^- \\ \thlf^- \end{array} \right)$ 
& $\left( \begin{array} {c} \quad-\quad   1.067 \;\; 1.128 \\ 1.539 \;\; 1.388 \;\; 1.707 \end{array} \right)$ \vspace{2mm} \\ 

$1^{3}P_2\otimes 1P_\frac{3}{2}$              & $(\hlf^+,\thlf^+,\fhlf^+,\shlf^+)$           
& $1^{3}P_2\otimes 1S_\frac{1}{2}$ & $\left( \begin{array} {c} \thlf^- \\ \fhlf^- \end{array} \right)$ 
& $\left( \begin{array} {c} 2.830 \;\; 0.626 \;\; 1.913 \;\; 1.800 \\ 
                            1.689 \;\; 1.369 \;\; 1.560 \;\; 2.057 \end{array} \right)$ \vspace{2mm} \\ 

 & & & & \large{$A^1_{\hlf, \hlf}$} \vspace{1mm} \\ \cline{5-5}

$1^{1}S_0\otimes 2S_\frac{1}{2}$              & $(\hlf^+)$           
& $1^{1}S_0\otimes 1S_\frac{1}{2}$ & $(\hlf^+)$ & (0.133)  \vspace{1mm} \\ 

 & & & & \large{$A^1_{\thlf, \hlf}$} \vspace{1mm} \\ \cline{5-5}

$1^{1}S_0\otimes 1D_\frac{3}{2}$              & $(\thlf^+)$           
&$1^{1}S_0\otimes 1S_\frac{1}{2}$ & $(\hlf^+)$ & (0.379)  \vspace{1mm} \\ 

 & & & & \large{$A^3_{\fhlf, \hlf}$} \vspace{1mm} \\ \cline{5-5}

$1^{1}S_0\otimes 1D_\frac{5}{2}$              & $(\fhlf^+)$           
& $1^{1}S_0\otimes 1S_\frac{1}{2}$ & $(\hlf^+)$ & (0.992)  \vspace{1mm} \\ \hline

\toprule
\end{tabular}

\end{table}
\end{center}
\begin{center}
\begin{table}[h]
\caption{Same as Table \ref{NoPStripwidths} but using wave functions from the spinless RP model including the $\emph{ {\bf l}}_\lambda \cdot {\bf s}_l$ interaction.
\label{MSOtripwidths}}
\vspace{2mm}
\resizebox{7.07in}{!} {
\begin{tabular}{c c c c c c c}
\hline \toprule
\multirow{2}{*}{Parent Multiplet} & \multirow{2}{*}{$J_a^{P_a}$} & \multirow{2}{*}{Daughter Multiplet} & \multirow{2}{*}{$\qquad J_b^{P_b} \qquad$} & \multicolumn{3}{c}{HDS Amplitude $A^l_{J_l, J^\prime_l}$ (GeV$^{-l-\hlf}$)} \\ 
 & && & \large{$\Xi_{cc}$} & \large{$\Xi_{bc}$} & \large{$\Xi_{bb}$} \\ \hline

 & & & & \multicolumn{3}{c}{\large{$A^0_{\hlf, \hlf}$}} \vspace{1mm} \\ \cline{5-7} 

$1^{3}S_1\otimes 1P_\frac{1}{2}$              & $(\hlf^-,\thlf^-)$           
& $1^{3}S_1\otimes 1S_\frac{1}{2}$ & $\left( \begin{array} {c} \hlf^+ \\ \thlf^+ \end{array} \right)$ 
& $\left( \begin{array} {c} 0.807 \;\;  \quad-\quad  \\  \quad-\quad  \;\; 0.810 \end{array} \right)$  
& $\left( \begin{array} {c} 0.805 \;\;  \quad-\quad  \\  \quad-\quad  \;\; 0.802 \end{array} \right)$  
& $\left( \begin{array} {c} 0.795 \;\;  \quad-\quad  \\  \quad-\quad  \;\; 0.796 \end{array} \right)$ \vspace{2mm} \\  

$1^{1}P_1\otimes 1P_\frac{1}{2}$              & $(\hlf^+,\thlf^+)$           
& $1^{1}P_1\otimes 1S_\frac{1}{2}$ & $\left( \begin{array} {c} \hlf^- \\ \thlf^- \end{array} \right)$
& $\left( \begin{array} {c} 0.601 \;\;  \quad-\quad  \\  \quad-\quad  \;\; 0.713 \end{array} \right)$  
& $\left( \begin{array} {c} 0.612 \;\;  \quad-\quad  \\  \quad-\quad  \;\; 0.705 \end{array} \right)$  
& $\left( \begin{array} {c} 0.601 \;\;  \quad-\quad  \\  \quad-\quad  \;\; 0.687 \end{array} \right)$ \vspace{2mm} \\  

 & & & & \multicolumn{3}{c}{\large{$A^2_{\thlf, \hlf}$}} \vspace{1mm} \\ \cline{5-7} 

$1^{3}S_1\otimes 1P_\frac{3}{2}$              & $(\hlf^-,\thlf^-,\fhlf^-)$           
& $1^{3}S_1\otimes 1S_\frac{1}{2}$ & $\left( \begin{array} {c} \hlf^+ \\ \thlf^+ \end{array} \right)$ 
& $\left( \begin{array} {c} \quad-\quad   2.097 \;\; 2.097 \\ 2.096 \;\; 2.096 \;\; 2.096 \end{array} \right)$  
& $\left( \begin{array} {c} \quad-\quad   2.040 \;\; 2.040 \\ 2.041 \;\; 2.041 \;\; 2.041 \end{array} \right)$  
& $\left( \begin{array} {c} \quad-\quad   1.998 \;\; 1.998 \\ 1.997 \;\; 1.997 \;\; 1.997 \end{array} \right)$ \vspace{2mm} \\ 

$1^{1}P_1\otimes 1P_\frac{3}{2}$              & $(\hlf^+,\thlf^+,\fhlf^+)$           
& $1^{1}P_1\otimes 1S_\frac{1}{2}$ & $\left( \begin{array} {c} \hlf^- \\ \thlf^- \end{array} \right)$
& $\left( \begin{array} {c} \quad-\quad   1.867 \;\; 2.155 \\ 1.738 \;\; 1.867 \;\; 2.155 \end{array} \right)$  
& $\left( \begin{array} {c} \quad-\quad   1.803 \;\; 2.096 \\ 1.667 \;\; 1.803 \;\; 2.096 \end{array} \right)$  
& $\left( \begin{array} {c} \quad-\quad   1.732 \;\; 2.037 \\ 1.594 \;\; 1.732 \;\; 2.037 \end{array} \right)$ \vspace{2mm} \\ 

 & & & & \multicolumn{3}{c}{\large{$A^1_{\hlf, \hlf}$}} \vspace{1mm} \\ \cline{5-7} 

$1^{3}S_1\otimes 2S_\frac{1}{2}$              & $(\hlf^+,\thlf^+)$           
& $1^{3}S_1\otimes 1S_\frac{1}{2}$ & $\left( \begin{array} {c} \hlf^+ \\ \thlf^+ \end{array} \right)$
& $\left( \begin{array} {c} 0.137 \;\;  0.143  \\  0.137  \;\; 0.142 \end{array} \right)$  
& $\left( \begin{array} {c} 0.120 \;\;  0.114  \\  0.120  \;\; 0.114 \end{array} \right)$  
& $\left( \begin{array} {c} 0.101 \;\;  0.104  \\  0.101  \;\; 0.104 \end{array} \right)$ \vspace{2mm} \\  

 & & & & \multicolumn{3}{c}{\large{$A^1_{\thlf, \hlf}$}} \vspace{1mm} \\ \cline{5-7} 

$1^{3}S_1\otimes 1D_\frac{3}{2}$              & $(\hlf^+,\thlf^+,\fhlf^+)$           
& $1^{3}S_1\otimes 1S_\frac{1}{2}$ & $\left( \begin{array} {c} \hlf^+ \\ \thlf^+ \end{array} \right)$ 
& $\left( \begin{array} {c} 0.476 \;\; 0.478, \quad-\quad \\ 0.478 \;\; 0.480 \;\; 0.473 \end{array} \right)$  
& $\left( \begin{array} {c} 0.442 \;\; 0.440, \quad-\quad \\ 0.440 \;\; 0.438 \;\; 0.431 \end{array} \right)$  
& $\left( \begin{array} {c} 0.411 \;\; 0.412, \quad-\quad \\ 0.412 \;\; 0.413 \;\; 0.407 \end{array} \right)$ \vspace{2mm} \\

 & & & & \multicolumn{3}{c}{\large{$A^3_{\fhlf, \hlf}$}} \vspace{1mm} \\ \cline{5-7}  

$1^{3}S_1\otimes 1D_\frac{5}{2}$              & $(\thlf^+,\fhlf^+,\shlf^+)$           
& $1^{3}S_1\otimes 1S_\frac{1}{2}$ & $\left( \begin{array} {c} \hlf^+ \\ \thlf^+ \end{array} \right)$  
& $\left( \begin{array} {c} \quad-\quad   1.061 \;\; 1.034 \\ 1.034 \;\; 1.060 \;\; 1.034 \end{array} \right)$  
& $\left( \begin{array} {c} \quad-\quad   0.977 \;\; 0.946 \\ 0.947 \;\; 0.979 \;\; 0.947 \end{array} \right)$  
& $\left( \begin{array} {c} \quad-\quad   0.921 \;\; 0.880 \\ 0.880 \;\; 0.921 \;\; 0.880 \end{array} \right)$ \vspace{2mm} \\  \hline

\toprule
\end{tabular}
}
\end{table}
\end{center}
\begin{center}
\begin{table}[h]
\caption{Same as Table \ref{NoPSsingwidths} but using wave functions from the spinless RP model including the $\emph{ {\bf l}}_\lambda \cdot {\bf s}_l$ interaction.
\label{MSOsingwidths}}
\vspace{2mm}
\begin{tabular}{c c c c c}
\hline \toprule
\multirow{2}{*}{Parent Multiplet} & \multirow{2}{*}{$J_a^{P_a}$} & \multirow{2}{*}{Daughter Multiplet} & \multirow{2}{*}{$\qquad J_b^{P_b} \qquad$} & HDS Amplitude $A^l_{J_l, J^\prime_l}$ (GeV$^{-l-\hlf}$) \\ 
 & && &  \large{$\Xi^\prime_{bc}$}  \\ \hline

 & & & & \large{$A^0_{\hlf, \hlf}$} \vspace{1mm} \\ \cline{5-5}

$1^{1}S_0\otimes 1P_\frac{1}{2}$              & $(\hlf^-)$           
&$1^{1}S_0\otimes 1S_\frac{1}{2}$ & $(\hlf^+)$  & (0.800) \vspace{1mm} \\ 

$1^{3}P_0\otimes 1P_\frac{1}{2}$              & $(\hlf^+)$           
& $1^{3}P_0\otimes 1S_\frac{1}{2}$ & $(\hlf^-)$ & (0.236) \vspace{1mm} \\ 

$1^{3}P_1\otimes 1P_\frac{1}{2}$              & $(\hlf^+,\thlf^+)$           
& $1^{3}P_1\otimes 1S_\frac{1}{2}$ & $\left( \begin{array} {c} \hlf^- \\ \thlf^- \end{array} \right)$
& $\left( \begin{array} {c} 0.205 \;\;  \quad-\quad  \\  \quad-\quad  \;\; 0.471 \end{array} \right)$ \vspace{2mm} \\  

$1^{3}P_2\otimes 1P_\frac{1}{2}$              & $(\thlf^+,\fhlf^+)$           
& $1^{3}P_2\otimes 1S_\frac{1}{2}$ & $\left( \begin{array} {c} \thlf^- \\ \fhlf^- \end{array} \right)$
& $\left( \begin{array} {c} 0.409 \;\;  \quad-\quad  \\  \quad-\quad  \;\; 0.744 \end{array} \right)$ \vspace{2mm} \\  

 & & & & \large{$A^2_{\thlf, \hlf}$} \vspace{1mm} \\ \cline{5-5}

$1^{1}S_0\otimes 1P_\frac{3}{2}$              & $(\thlf^-)$           
&$1^{1}S_0\otimes 1S_\frac{1}{2}$ & $(\hlf^+)$ & (2.050) \vspace{1mm} \\ 

$1^{3}P_0\otimes 1P_\frac{3}{2}$              & $(\thlf^+)$           
&$1^{3}P_0\otimes 1S_\frac{1}{2}$ & $(\hlf^-)$ & (1.569) \vspace{1mm} \\ 

$1^{3}P_1\otimes 1P_\frac{3}{2}$              & $(\hlf^+,\thlf^+,\fhlf^+)$           
& $1^{3}P_1\otimes 1S_\frac{1}{2}$ & $\left( \begin{array} {c} \hlf^- \\ \thlf^- \end{array} \right)$
& $\left( \begin{array} {c} \quad-\quad   1.841 \;\; 1.432 \\ 1.629 \;\; 0.823 \;\; 2.013 \end{array} \right)$ \vspace{2mm} \\ 

$1^{3}P_2\otimes 1P_\frac{3}{2}$              & $(\hlf^+,\thlf^+,\fhlf^+,\shlf^+)$           
& $1^{3}P_2\otimes 1S_\frac{1}{2}$ & $\left( \begin{array} {c} \thlf^- \\ \fhlf^- \end{array} \right)$
& $\left( \begin{array} {c} 2.908 \;\; 0.628 \;\; 1.924 \;\; 1.914 \\ 
                            1.585 \;\; 1.370 \;\; 1.573 \;\; 2.086 \end{array} \right)$ \vspace{2mm} \\ 

 & & & & \large{$A^1_{\hlf, \hlf}$} \vspace{1mm} \\ \cline{5-5}

$1^{1}S_0\otimes 2S_\frac{1}{2}$              & $(\hlf^+)$           
&$1^{1}S_0\otimes 1S_\frac{1}{2}$ & $(\hlf^+)$ & (0.113)  \vspace{1mm} \\ 

 & & & & \large{$A^1_{\thlf, \hlf}$} \vspace{1mm} \\  \cline{5-5}

$1^{1}S_0\otimes 1D_\frac{3}{2}$              & $(\thlf^+)$           
&$1^{1}S_0\otimes 1S_\frac{1}{2}$ & $(\hlf^+)$ & (0.431)  \vspace{1mm} \\ 

 & & & & \large{$A^3_{\fhlf, \hlf}$} \vspace{1mm} \\  \cline{5-5}

$1^{1}S_0\otimes 1D_\frac{5}{2}$              & $(\fhlf^+)$           
&$1^{1}S_0\otimes 1S_\frac{1}{2}$ & $(\hlf^+)$ & (0.986)  \vspace{1mm} \\ \hline

\toprule
\end{tabular}

\end{table}
\end{center}

\subsubsection{Excitation Symmetry}

In Tables \ref{NoPStripwidths} and \ref{NoPSsingwidths} the amplitudes are expected to depend on the quantum numbers of the light degrees of freedom, but should  be independent of the quantum numbers of the diquark. This means that the amplitudes in the first two rows of Table \ref{NoPStripwidths}, for the decays of a particular flavor of baryon, should all be equal, as they all describe transitions between states in which the light degrees of freedom change from $1P_{\hlf}$ to $1S_{\hlf}$.  In the same way, the numbers in rows three and four should be the same. Similarly, in Table \ref{NoPSsingwidths} all of the amplitudes in the first four rows should be the same, and the amplitudes in rows five to eight should also be identical.  For the most part, these expectations are realized, but there are some significant departures.   As one might expect from the discussion of mixing in the previous section, the worst disagreement between this model and the expectations of excitation symmetry comes from decays of flavor-singlet states (Table \ref{NoPSsingwidths}).  Amplitudes for the $1^{3}P_2\otimes 1P_\frac{3}{2} \rightarrow 1^{3}P_2\otimes 1S_\frac{1}{2}$ transition differ from amplitudes for the $1^{3}P_1\otimes 1P_\frac{3}{2} \rightarrow 1^{3}P_1\otimes 1S_\frac{1}{2}$ transitions significantly. 

The excitation symmetry and the decoupling of the total angular momentum of the heavy diquark suggest that  the strong decay amplitudes will also be independent of the statistics of the heavy diquark.  Therefore, the HDS amplitudes for  particular light-component transition in Table \ref{NoPSsingwidths} should be equal to the corresponding light-component transition amplitudes in the $\Xi_{bc}$ column of Table \ref{NoPStripwidths}. For example, amplitudes for the transitions from the $1P_\hlf$ to $1S_\hlf$ in Table \ref{NoPStripwidths} (rows 1 and 2) should be equal to those for the same transitions in Table~\ref{NoPSsingwidths} (rows 1 to 4).  As with the diquark excitation symmetry, this works well in some cases, not so well in others, and the causes of these deviations have been discussed above: various mixings in the wave functions, and zeroes in the amplitudes for the decays from states in which the light degrees of freedom are radially excited. 

Tables \ref{MSOtripwidths} and \ref{MSOsingwidths} show the results obtained using the $\mu$SO model.  For the flavor-triplet states, these results are slightly more consistent with the expectations of the diquark excitation symmetry. As with the full RP model, there are deviations from this symmetry.  For flavor-singlet states, the deviations are larger. These deviations are primarily due to pairwise mixing as was discussed in the previous section.  The symmetry between flavor-singlet and flavor-triplet states can be seen by comparing the HDS amplitudes in Table \ref{MSOsingwidths} to the corresponding amplitudes in the $\Xi_{bc}$ column of Table \ref{MSOtripwidths}. 

\subsubsection{Diquark Flavor Symmetry}

It is expected that as the mass of the heavy diquark becomes very large the wave function of the light degrees of freedom becomes independent of the heavy diquark mass.  As a result, decays which involve resonances which are eigenstates of HDS should respect this symmetry if the heavy diquark is sufficiently massive.  Thus, all of the numbers in any row of Tables \ref{NoPStripwidths}and \ref{NoPSsingwidths} should be the same. However, amplitudes calculated in the $^3P_0$ model depend on the mass of the states involved in a somewhat complicated way through the center of mass three-momentum $k_0$.  We have accounted for this partially by dividing off a factor of $k_0^l$ from the decay amplitude, but this does not remove all of the $k_0$ dependence. The amplitudes shown in the tables decrease very  slightly as the mass of the heavy diquark increases, but the numbers in columns 5, 6 and 7 of these two  tables  are nevertheless consistent with the expectations of diquark flavor symmetry.  In this case, results for the decays of states in which the light degrees of freedom are in a $D$ wave, or are radially excited, also exhibit the flavor symmetry. For these states, the symmetry is much better exhibited in the $\mu$SO version of the model (Tables \ref{MSOtripwidths} and \ref{MSOsingwidths}). As with all of the other symmetries we have discussed, the results for this symmetry are subject to the same kinds of deviations induced by mixings in the wave functions, etc.

Again, we emphasize that this model does not assume a pointlike heavy diquark and allows excitations of the heavy degrees of freedom to mix with excitations of the light degrees of freedom.  Furthermore, the part of the wave functions corresponding to the light degrees of freedom may, in principle, differ significantly between $\Xi_{cc}$ baryons and $\Xi_{bb}$ baryons.  The similarity between these particles embodied in these results is an emergent one.  This is in contrast to an interpretation of results from NRQCD \cite{Thacker91}, where the heavy quark flavor symmetry is broken at lowest order in the heavy quark expansion for the heavy diquark subsystem.  The results of the present model indicate that the physics of the heavy diquark decouple sufficiently from the physics of the light degrees of freedom that an approximate heavy diquark flavor symmetry emerges for the {\textit heavy}-{\textit light} subsystem.  This observation also seems to hold for the mass spectrum of these baryons \cite{eakinsroberts}.

\subsection{Strange Diquarks}

In many quark models, the strange quark's mass is similar in magnitude to $\Lambda_{\rm QCD}$. This should be much too light for this quark to be considered heavy, but features of decay widths emerge that are very easily understood if this odd approximation is made.  If the strange quark is treated as heavy, the flavor multiplets of section~\ref{flavormultiplets} need to be modified. Instead of the broken SU(2) of that section, we now consider a (badly) broken ${\text SU(3)}_{\text heavy}$ flavor symmetry.  There will now be a sextet of symmetric states
\begin{equation}
 \Xi=ssq,\,\,\,\,\,\ \Xi_{cc}=ccq,\,\,\,\,\,\Xi_{bb}=bbq,\,\,\,\,\Xi_{c}=\frac{1}{\sqrt{2}}\left(sc+cs\right)q,\,\,\,\,\Xi_{b}=\frac{1}{\sqrt{2}}\left(sb+bs\right)q,\,\,\,\,\Xi_{bc}=\frac{1}{\sqrt{2}}\left(cb+bc\right)q,
\end{equation}
and an antitriplet of antisymmetric states
\begin{equation}
\Xi^\prime_{c}=\frac{1}{\sqrt{2}}\left(sc-cs\right)q,\,\,\,\,\Xi^\prime_{b}=\frac{1}{\sqrt{2}}\left(sb-bs\right)q,\,\,\,\,\Xi^\prime_{bc}=\frac{1}{\sqrt{2}}\left(cb-bc\right)q.
\end{equation}
For these states, there are only two choices for the light quark, $q$, namely $u$ and $d$.  These multiplets should not be confused with the multiplets in which singly-heavy baryons are placed, with the strange quark taken as one of the {\em light} triplet of quarks. The ${\text SU(3)}_{\text light}$ sextet then contains the states
\begin{equation}
 \Sigma_Q=uuQ,\,\, ddQ,\,\,\frac{1}{\sqrt{2}}\left(ud+du\right)Q;\,\,\,\Xi^\prime_{Q}=\frac{1}{\sqrt{2}}\left(su+us\right)Q, \,\,\frac{1}{\sqrt{2}}\left(sd+ds\right)Q;\,\,\,\,\Omega_{Q}=ssQ,
\end{equation}
and the corresponding antitriplet consists of
\begin{equation}
\Lambda_Q=\frac{1}{\sqrt{2}}\left(ud-du\right)Q,\,\,\,\,\Xi_{Q}=\frac{1}{\sqrt{2}}\left(us-su\right)Q,\,\,\frac{1}{\sqrt{2}}\left(ds-sd)\right)Q. 
\end{equation}
In these multiplets, the heavy quark $Q$ may be either $b$ or $c$.

In the following subsections we examine the consequences of treating the strange quark as a heavy quark in doubly heavy systems. The symmetries we have discussed in the preceding sections should not be expected to work well when applied to baryons in which the strange quark is treated as heavy. Nevertheless, examining such a limit may provide clues to the understanding of the spectroscopy of such baryons, particularly the `light' $\Xi$s. It may also provide some information on how badly the symmetries get broken as the masses of the heavy quarks decrease. 

\subsubsection{Angular Momentum Multiplets and Decay Widths}

The pion-emission decay widths of the flavor-sextet $\Xi$, $\Xi_{c}$, and $\Xi_{b}$ states are shown in Table \ref{Chitripwidths}. All decay widths shown in that table are for decays to the ground state $n_d^{2s_d+1}(l_\rho)_{J_d}\otimes n_\lambda(l_\lambda)_{J_l}=1^{3}S_1\otimes 1S_\frac{1}{2}$. The analogous decays of the flavor-antitriplet $\Xi_{c}^\prime$ and $\Xi_{b}^\prime$ to the ground-state $n_d^{2s_d+1}(l_\rho)_{J_d}\otimes n_\lambda(l_\lambda)_{J_l}=1^{1}S_0\otimes 1S_\frac{1}{2}$ are shown in Table \ref{Chisingwidths}.  The format of the tables is explained in section~\ref{DHBwidths}.  We should caution that the masses and decay widths presented here ignore mixing of $\Xi_c$ and $\Xi^\prime_c$ states as well as the $\Xi_b$ and $\Xi^\prime_b$ states, but the original model allowed this mixing, and model parameters were fit to experiment with this mixing included.  Therefore, the masses presented here will be different from those originally published \cite{RobertsPervin}. 

\begin{center}
\begin{table}[h]
\caption{Decay widths for pion-emission from the parent multiplet listed to the ground state $n_d^{2s_d+1}(l_\rho)_{J_d}\otimes n_\lambda(l_\lambda)_{J_l}=1^{3}S_1\otimes 1S_\frac{1}{2}$ doublet for strange flavor-sextet states.  The format of the table is explained in section~\ref{DHBwidths}.  When available the experimental masses are listed below the masses obtained from the RP model \cite{RobertsPervin}.
\label{Chitripwidths}}
\vspace{2mm}
\resizebox{7.07in}{!} {
\begin{tabular}{c c c c c c c}
\hline \toprule

Parent Multiplet, $J_a^{P_a}$   & \multicolumn{6}{c }{Masses (GeV) and Widths (MeV)} \\ 
$n_d^{2s_d+1}(l_\rho)_{J_d}\otimes  n_\lambda(l_\lambda)_{J_l}$ & \multicolumn{2}{c}{\large{$\Xi$}} & \multicolumn{2}{c}{\large{$\Xi_{c}$}}& \multicolumn{2}{c}{\large{$\Xi_{b}$}} \\ \cmidrule{2-3} \cmidrule{4-5} \cmidrule{6-7}
& Mass (GeV) & Width (MeV)  &Mass (GeV) & Width (MeV)& Mass (GeV) & Width (MeV)\\\hline

$\begin{array}{c}1^{1}P_1 \otimes  1S_\frac{1}{2}\\  (\hlf^-,\thlf^-)\end{array}$ &$\begin{array}{c} (1.725, 1.759) \\(1.690,1.823)\end{array}$        
& $\left( \begin{array} {c} 13.88 \:\: 0.402 \\ 0.002 \:\: 26.32 \end{array} \right)$  & (2.861, 2.860)
& $\left( \begin{array} {c} 5.245 \:\: 0.042 \\ 0.041 \:\: 3.435 \end{array} \right)$  & (6.207, 6.206)
& $\left( \begin{array} {c} 4.128 \:\: 0.023 \\ 0.045 \:\: 6.371 \end{array} \right)$ \vspace{2mm} \\ 

$\begin{array}{c}2^{3}S_1 \otimes  1S_\frac{1}{2} \\ (\hlf^+,\thlf^+)\end{array}$ & (1.891, 2.021)        
& $\left( \begin{array} {c} 1.258 \:\: 3.436 \\ 18.92 \:\: 0.060 \end{array} \right)$  & (3.030, 3.096)
& $\left( \begin{array} {c} 0.284 \:\: 0.354 \\ 4.691 \:\: 0.000 \end{array} \right)$  & (6.377, 6.424)
& $\left( \begin{array} {c} 0.402 \:\: 0.166 \\ 5.251 \:\: 0.000 \end{array} \right)$ \vspace{2mm} \\  

$\begin{array}{c}1^{3}D_1\otimes  1S_\frac{1}{2} \\   (\hlf^+,\thlf^+)\end{array}$& (2.055, 1.934\footnotemark[1])
& $\left( \begin{array} {c} 0.030 \:\: 0.546 \\ 0.935 \:\: 0.103 \end{array} \right)$  & (3.119, 3.053\footnotemark[1])
& $\left( \begin{array} {c} 0.030 \:\: 0.255 \\ 0.119 \:\: 0.080 \end{array} \right)$  & (6.450, 6.408\footnotemark[1])
& $\left( \begin{array} {c} 0.035 \:\: 0.289 \\ 0.052 \:\: 0.196 \end{array} \right)$ \vspace{2mm} \\ 

$\begin{array}{c}1^{3}D_2\otimes  1S_\frac{1}{2}  \\  (\thlf^+,\fhlf^+)\end{array}$ & (2.032\footnotemark[1], 2.025)        
& $\left( \begin{array} {c} 0.027 \:\: 0.000 \\ 0.160 \:\: 0.031 \end{array} \right)$  & (3.113\footnotemark[1], 3.109)
& $\left( \begin{array} {c} 0.096 \:\: 0.001 \\ 0.059 \:\: 0.028 \end{array} \right)$  & (6.446\footnotemark[1], 6.443)
& $\left( \begin{array} {c} 0.113 \:\: 0.001 \\ 0.025 \:\: 0.020 \end{array} \right)$ \vspace{2mm} \\          

$\begin{array}{c}1^{3}D_3\otimes  1S_\frac{1}{2}\\  (\fhlf^+,\shlf^+)\end{array}$    & (1.936, 2.035)      
& $\left( \begin{array} {c} 0.015 \:\: 0.000 \\ 0.993 \:\: 0.000 \end{array} \right)$  & (3.053, 3.115)
& $\left( \begin{array} {c} 0.017 \:\: 0.005 \\ 0.595 \:\: 0.001 \end{array} \right)$  & (6.409, 6.447)
& $\left( \begin{array} {c} 0.016 \:\: 0.003 \\ 0.959 \:\: 0.001 \end{array} \right)$ \vspace{2mm} \\       

$\begin{array}{c}1^{3}S_1\otimes  1P_\frac{1}{2} \\  (\hlf^-,\thlf^-)\end{array}$  & (1.826\footnotemark[3], 1.826\footnotemark[4])        
& $\left( \begin{array} {c} 201.2 \:\: 6.468 \\ 1.858 \:\: 131.0 \end{array} \right)$ & (2.976, 2.952\footnotemark[4]) 
& $\left( \begin{array} {c} 249.2 \:\: 8.877 \\ 0.382 \:\: 230.5 \end{array} \right)$  & (6.311, 6.287\footnotemark[4])
& $\left( \begin{array} {c} 351.8 \:\: 4.113 \\ 0.604 \:\: 230.5 \end{array} \right)$ \vspace{2mm} \\ 

$\begin{array}{c}1^{3}S_1\otimes  1P_\frac{3}{2} \\  (\hlf^-,\thlf^-,\fhlf^-)\end{array}$    & (1.811\footnotemark[3], 1.849\footnotemark[4], 1.883)      
& $\left( \begin{array} {c} 133.9 \:\: 14.15 \:\: 74.89 \\ 2.883 \:\: 135.8 \:\: 9.347 \end{array} \right)$  & (2.952, 2.963\footnotemark[4], 2.970)
& $\left( \begin{array} {c} 5.320 \:\: 12.83 \:\: 63.43 \\ 10.16 \:\: 84.52 \:\: 6.951 \end{array} \right)$  & (6.287, 6.302\footnotemark[4], 6.307)
& $\left( \begin{array} {c} 12.59 \:\: 5.371 \:\: 29.31 \\ 9.654 \:\: 109.5 \:\: 6.967 \end{array} \right)$ \vspace{2mm} \\           

$\begin{array}{c}1^{1}P_1\otimes  1P_\frac{1}{2} \\  (\hlf^+,\thlf^+)\end{array}$   & (2.108\footnotemark[2], 2.106)
& $\left( \begin{array} {c} 0.607 \:\: 2.787 \\ 0.000 \:\: 1.553 \end{array} \right)$  & (3.206\footnotemark[2], 3.195) 
& $\left( \begin{array} {c} 0.189 \:\: 0.859 \\ 0.165 \:\: 0.667 \end{array} \right)$  & (6.523\footnotemark[2], 6.531)
& $\left( \begin{array} {c} 0.127 \:\: 0.905 \\ 0.067 \:\: 1.352 \end{array} \right)$ \vspace{2mm} \\          

$\begin{array}{c}1^{1}P_1\otimes  1P_\frac{3}{2}  \\  (\hlf^+,\thlf^+,\fhlf^+)\end{array}$ & (2.129\footnotemark[2]\footnotemark[5], 2.115, 2.137\footnotemark[6])                   
& $\left( \begin{array} {c} 0.094 \:\: 0.013 \:\: 1.711 \\ 4.506 \:\: 0.053 \:\: 66.75 \end{array} \right)$  & (3.162\footnotemark[2]\footnotemark[5], 3.206, 3.195)
& $\left( \begin{array} {c} 0.686 \:\: 0.044 \:\: 0.060 \\ 30.40 \:\: 0.012 \:\: 0.918 \end{array} \right)$ & (6.508\footnotemark[2], 6.542, 6.530)
& $\left( \begin{array} {c} 0.574 \:\: 0.025 \:\: 0.005 \\ 13.14 \:\: 0.012 \:\: 1.986 \end{array} \right)$ \vspace{2mm} \\           

$\begin{array}{c}1^{3}S_1\otimes  2S_\frac{1}{2} \\ (\hlf^+,\thlf^+)\end{array}$  & (2.014\footnotemark[5], 2.144)            
& $\left( \begin{array} {c} 6.375 \:\: 4.115 \\ 57.84 \:\: 1.325 \end{array} \right)$  & (3.202\footnotemark[5], 3.256)
& $\left( \begin{array} {c} 0.000 \:\: 12.35 \\ 20.35 \:\: 2.953 \end{array} \right)$   & (6.538, 6.579)
& $\left( \begin{array} {c} 0.514 \:\: 1.390 \\ 24.46 \:\: 4.491 \end{array} \right)$ \vspace{2mm} \\         

$\begin{array}{c}1^{3}S_1\otimes  1D_\frac{3}{2} \\ (\hlf^+,\thlf^+,\fhlf^+)\end{array}$  & (2.185, 2.149, 2.143\footnotemark[6])
& $\left( \begin{array} {c} 122.0 \:\: 84.60 \:\: 7.851 \\ 13.05 \:\: 64.01 \:\: 33.60 \end{array} \right)$  & (3.277, 3.274, 3.267)
& $\left( \begin{array} {c} 73.71 \:\: 48.77 \:\: 6.940 \\ 14.50 \:\: 40.11 \:\: 201.7 \end{array} \right)$  & (6.609,  6.606, 6.596)
& $\left( \begin{array} {c} 110.6 \:\: 66.91 \:\: 9.802 \\ 15.95 \:\: 30.73 \:\: 220.5 \end{array} \right)$ \vspace{2mm} \\            

$\begin{array}{c}1^{3}S_1\otimes  1D_\frac{5}{2}  \\  (\thlf^+,\fhlf^+,\shlf^+)\end{array}$  & (2.142, 2.148\footnotemark[6], 2.148) 

& $\left( \begin{array} {c} 0.021 \:\: 22.58 \:\: 75.49 
                         \\ 47.22 \:\: 81.48 \:\: 23.92 \end{array} \right)$  & (3.268, 3.270, 3.275)
& $\left( \begin{array} {c} 1.279 \:\: 35.57 \:\: 102.9 
                         \\ 97.70 \:\: 5.340 \:\: 33.61 \end{array} \right)$  & (6.598,  6.600, 6.607)
& $\left( \begin{array} {c} 4.558 \:\: 23.58 \:\: 82.85 
                         \\ 120.8 \:\: 0.045 \:\: 38.28 \end{array} \right)$ \vspace{2mm} \\\hline           
\toprule
\footnotetext[1]{$^{\text{bcdef}}$These states mix significantly.}
\end{tabular}}
\end{table}
\end{center}

\begin{center}
\begin{table}[h]
\caption{Decay widths for pion-emission from the parent multiplet listed to the ground state $n_d^{2s_d+1}(l_\rho)_{J_d}\otimes n_\lambda(l_\lambda)_{J_l}=1^{1}S_0\otimes 1S_\frac{1}{2}$ singlet for strange flavor-antitriplet states.  The format of the table is explained in section~\ref{DHBwidths}.
\label{Chisingwidths}}
\vspace{2mm}
\renewcommand{\arraystretch}{1.0}
\begin{tabular}{c c c c c }
\hline \toprule
Parent Multiplet, $J_a^{P_a}$   & \multicolumn{2}{c}{\large{$\Xi_{c}^\prime$}} & \multicolumn{2}{c}{\large{$\Xi_{b}^\prime$}}\\ 
$n_d^{2s_d+1}(l_\rho)_{J_d}\otimes  n_\lambda(l_\lambda)_{J_l}$ 
& Mass (GeV) & Width (MeV)  &Mass (GeV) & Width (MeV)\\\hline

$\begin{array}{c}1^{3}P_0\otimes 1S_\frac{1}{2}\\ (\hlf^-)\end{array}$  & (2.907\footnotemark[1])   & (8.663)   & (6.236\footnotemark[1])      & (1.974) \vspace{1mm}  \\   

$\begin{array}{c}1^{3}P_1\otimes 1S_\frac{1}{2}\\(\hlf^-,\thlf^-)\end{array}$  &  (2.814\footnotemark[1], 2.899)  & (11.893  0.001)  & (6.173\footnotemark[1], 6.228)& (8.669  0.000) \vspace{1mm} \\ 

$\begin{array}{c}1^{3}P_2\otimes 1S_\frac{1}{2}\\(\thlf^-,\fhlf^-)\end{array}$ &  (2.816, 2.890)  & (0.064  0.012)  & (6.173, 6.227)& (0.038  0.005) \vspace{1mm} \\ 

$\begin{array}{c}2^{1}S_0\otimes 1S_\frac{1}{2}\\(\hlf^+)\end{array}$  & (3.032)  & (0.075)   & (6.390)    & (0.364) \vspace{1mm} \\ 

$\begin{array}{c}1^{1}D_2\otimes 1S_\frac{1}{2}\\(\thlf^+,\fhlf^+)\end{array}$  &  (3.091, 3.090)  & (0.549  0.060)  & (6.434, 6.432)& (0.524  0.057)  \vspace{1mm}\\

$\begin{array}{c}1^{1}S_0\otimes 1P_\frac{1}{2}\\(\hlf^-)\end{array}$ &  (2.927)   & (338.8)   & (6.285)    & (365.4)\vspace{1mm} \\ 

$\begin{array}{c}1^{1}S_0\otimes 1P_\frac{3}{2}\\(\thlf^-)\end{array}$ & (2.927)  & (21.29)     & (6.285)    & (22.22) \vspace{1mm}\\ 

$\begin{array}{c}1^{3}P_0\otimes 1P_\frac{1}{2}\\(\hlf^+)\end{array}$  &(3.204\footnotemark[2])  & (0.087)  & (6.540\footnotemark[2])       & (0.092) \vspace{1mm}\\ 

$\begin{array}{c}1^{3}P_0\otimes 1P_\frac{3}{2}\\(\thlf^+)\end{array}$ & (3.202\footnotemark[3]) & (0.761)   & (6.539\footnotemark[3])       & (0.518) \vspace{1mm}\\ 

$\begin{array}{c}1^{3}P_1\otimes 1P_\frac{1}{2}\\(\hlf^+,\thlf^+)\end{array}$  & (3.216\footnotemark[2], 3.195\footnotemark[3])  & (1.356  0.002)  & (6.547\footnotemark[2], 6.531\footnotemark[3])  & (0.089  0.008) \vspace{1mm}\\ 

$\begin{array}{c}1^{3}P_1\otimes 1P_\frac{3}{2}\\(\hlf^+,\thlf^+,\fhlf^+)\end{array}$ & (3.212, 3.178\footnotemark[3], 3.205\footnotemark[4])  & (0.131  1.406  0.027)  & (6.545, 6.514\footnotemark[3], 6.536\footnotemark[4]) & (0.221  1.721  0.001)  \vspace{1mm} \\ 

$\begin{array}{c}1^{3}P_2\otimes 1P_\frac{1}{2}\\(\thlf^+,\fhlf^+)\end{array}$  &  (3.209, 3.177\footnotemark[4])  & (0.193  0.313)   & (6.544, 6.513\footnotemark[4])& (0.000  0.374) \vspace{1mm} \\ 

$\begin{array}{c}1^{3}P_2\otimes 1P_\frac{3}{2}\\(\hlf^+,\thlf^+,\fhlf^+,\shlf^+)\end{array}$ &$\begin{array}{c}(3.116\footnotemark[2], 3.206\footnotemark[3],\\ 3.199\footnotemark[4], 3.203)\end{array}$ & $\begin{array}{c}(12.88  0.159\\  0.002  0.168)\end{array}$      &$\begin{array}{c}(6.462\footnotemark[2], 6.542\footnotemark[3], \\ 6.541\footnotemark[4], 6.539)\end{array}$       & $\begin{array}{c}(6.087  0.030\\  0.015  0.124)\end{array}$ \vspace{1mm}\\ 

$\begin{array}{c}1^{1}S_0\otimes 2S_\frac{1}{2}\\(\hlf^+)\end{array}$      & (3.219)  & (4.800)   & (6.553)      & (7.045) \vspace{1mm} \\ 

$\begin{array}{c}1^{1}S_0\otimes 1D_\frac{3}{2}\\(\thlf^+)\end{array}$  & (3.241)  & (133.0)   & (6.587)       & (132.6) \vspace{1mm} \\ 

 $\begin{array}{c}1^{1}S_0\otimes 1D_\frac{5}{2}\\(\fhlf^+)\end{array}$    &  (3.239)  & (89.75)  & (6.586)       & (102.7)  \vspace{.5mm} \\\hline

\toprule \footnotetext[1]{$^{\text{bcd}}$These states mix significantly.}

\end{tabular}
\end{table}
\end{center}

We showed in section~\ref{DHBwidths} that this model predicts a significant suppression of DHB pion decays to the ground state multiplet for parent baryons consisting of an excited heavy diquark.  Systems with strange diquarks show a similar suppression.  For flavor-sextet states, the largest of these `suppressed' decay modes occurs for the $\Xi$ for the $\fhlf^+$ member of the $1^{1}P_1\otimes 1P_\frac{3}{2}$ multiplet decaying to the $\thlf^+$ member of the ground state doublet.  This width is 67 MeV for the $\Xi$ but drops to 2 MeV for the $\Xi_{b}$.  For the $\Xi^\prime_{c}$ the largest of the suppressed decay modes is the decay of the $\hlf^+$ member of the $1^{3}P_2\otimes 1P_\frac{3}{2}$ multiplet to the ground state singlet (12.9 MeV).  The size of these widths is largely due to various mixings in the wave functions, which decrease as the mass of the diquark increases.

There are also decay modes in Table \ref{Chitripwidths} whose suppression can be understood by spin-counting arguments.  For instance, the decay width of the $\hlf^-$ member of the $1^{3}S_1\otimes 1P_\frac{1}{2}$ multiplet to the ground state $\thlf^+$ is only 1.858 MeV for the $\Xi$.  This particular decay can only occur in a D-wave because of the conservation of total angular momentum and parity, but the $6-j$ symbol in Eq.~(\ref{HDS}) vanishes for this partial wave. Therefore, this decay only occurs through mixing between the $1^{3}S_1\otimes 1P_\frac{1}{2}$ and $1^{3}S_1\otimes 1P_\frac{3}{2}$ multiplets. This is completely analogous to the discussion in section \ref{DHBwidths}, regarding the decays of the $\Xi_{cc}$, $\Xi_{bc}$ and $\Xi_{bb}$.

Experimentally, there are two known $\Xi$ resonances which could be interpreted as belonging to particular excited HDS multiplets, with some degree of certainty.  A comparison of experimentally measured masses of $\Xi$ baryons \cite{PDG} to quark model masses suggests that the $\Xi(1690)$ and the $\Xi(1820)$ form the $1^{1}P_1\otimes 1S_\frac{1}{2}$ $(\hlf^-,\thlf^-)$ multiplet.  The notion that these states consist primarily of strange diquark excitations is also supported by the (limited) measurements of the decay modes of these states.  It is clear that these states prefer to decay by kaon rather than pion-emission.  This fact is readily explained by the diquark selection rule (see section \ref{selectrule}).  The strange diquark is not a spectator in kaon decays, implying that these processes are not suppressed by the symmetry.  For the $\Xi(1690)$ the PDG estimates that the width of this state is $<30$ MeV \cite{PDG}, and there is also an indication that the pion decay channel is a small fration of this width ($\Gamma(\Xi \pi)/\Gamma(\Sigma \overline{K}) < 0.09$), giving an upper limit of 2.7 MeV for the pion-emission width of this state.  In comparison, the present model predicts a width of about 14 MeV for this decay mode.  For the $\Xi(1820)$ the measured pion decay width is on the order of one MeV, while the model predicts about 26 MeV.  The `large' pion decay widths obtained in the model are due in part to mixing with the $1^{3}S_1\otimes 1P_\frac{1}{2}$ multiplet whose S-wave pion decay widths are quite large.  Therefore, even a small mixing with this multiplet will produce a large enhancement of these suppressed widths.  

There is a similar indication of suppression of pion decays for the $\Xi(2030)$.  The parity of this resonance is unknown, but the spin has been determined to be $J \ge \fhlf$.  Comparing the mass predictions of the RP model to the measured mass yields two candidates for this resonance:  the 2025 MeV $\fhlf^+$ member of the $1^{1}D_2\otimes 1S_\frac{1}{2}$ multiplet and the 2035 MeV $\shlf^+$ member of the $1^{1}D_3\otimes 1S_\frac{1}{2}$ multiplet.  The predicted pion decay widths of these states are quite small in agreement with the measurements for the $\Xi(2030)$.  If the parity of this resonance is determined to be negative, the current form of the model predicts no negative-parity states that are this high in mass. However, there are more massive negative-parity states that will result if the expansion basis used in the model is made larger.  

Measurements of $\Xi_c$ decay widths can also be understood in terms of a $q-cs$ quark-diquark structure.  The $\Xi_c(2790)$ ($J^P=\hlf^-$) decays by pion-emission to the $\Xi^\prime_c$ with a width $<12$ MeV.  Based on the this state's mass, it is natural to assign it to an HDS multiplet with a diquark excitation \cite{eakinsroberts}.  In the RP model there are two candidates:  the flavor-antitriplet 2814 MeV state belonging to the $1^{3}P_1\otimes 1S_\frac{1}{2}$ HDS doublet and the flavor-antitriplet 2907 MeV state belonging to the $1^{3}P_0\otimes 1S_\frac{1}{2}$ HDS singlet.  We calculate the width of the $1^{3}P_1\otimes 1S_\frac{1}{2}$ $\hlf^-$ state to be 11.9 MeV and the width of the $1^{3}P_0\otimes 1S_\frac{1}{2}$ state to be 8.7 MeV (Table~\ref{Chisingwidths}).  Thus, either model state could be identified with this experimental state.

 The $\Xi_c(2815)$ ($J^P=\thlf^-$) decays with a width of $<3.5$ MeV by pion-emission to the $\Xi_c(2645)$ ($\thlf^+$) ground state.  It is natural to assign this state to the flavor-sextet $1^{1}P_1\otimes 1S_\frac{1}{2}$ HDS doublet.  The model predicts a mass of 2860 MeV and a width of 3.4 MeV for this state.  There are several other observed $\Xi$ and $\Xi_c$ resonances, but their decay widths and branching fractions have not been measured well enough to make a meaningful comparisons at this time.

One may also apply HQS to these decays by treating the strange quark as a light quark and working in the $SU(3)_{light}$ basis.  Such an investigation has already been carried out in the context of Heavy Hadron Chiral Perturbation Theory \cite{Cheng}.  In the $SU(3)_{light}$ basis, the ground states are a $J_l^{\pi_l}=0^+$, $J^P=\hlf^+$ singlet (the $\Xi_c$), and a $J_l^{\pi_l}=1^+$, $(\hlf^+,\thlf^+)$ doublet (the $\Xi^\prime_c$ and the $\Xi_c(2645)$).  The $\Xi_c(2790)$ and $\Xi_c(2815)$ make up a $(\hlf^-,\thlf^-)$ HQS doublet with $J_l^{\pi_l}=1^-$ strange/light degrees of freedom.  In this interpretation the $\Xi_c(2790)$ may only decay by an S-wave to the $\Xi^\prime_c$ because this is the only channel which conserves the total angular momentum of the light degrees of freedom.  Similarly, the $\Xi_c(2815)$ primarily decays to the $\Xi_c(2645)$ in an S-wave, although it may also decay by a D-wave transition to the $\Xi^\prime_c$ or the $\Xi_c(2645)$.  If the D-wave decay mode of the $\Xi_c(2815)$ is indeed negligible, then leading order HQS predicts that
\beq
\frac{\Gamma(\Xi_c(2790) \rightarrow \Xi^\prime_c \pi)}{\Gamma(\Xi_c(2815) \rightarrow \Xi_c(2645) \pi)}=1.54 \pm0.04.
\eeq
By contrast the Heavy Diquark Symmetry forbids both of these decays at lowest order, because they involve transitions of the diquark as discussed previously.  If this is the case, then these decays only occur through mixing of symmetry multiplets implying that a relationship between these two widths can only be obtained by considering higher order corrections to the symmetries we have proposed.

\subsubsection{HDS Spin-Counting}

Table \ref{XiSpinCount1+} shows ratios of decay amplitudes involving the ground state strange diquark.  These are calculated in three different variations of the RP model: the full model, the spinless model, and the $\mu$SO model, which is the spinless RP model appended with an $\rmb{l}_\lambda \cdot \rmb{s}_l$ interaction.  The appropriate spin-counting factors have been divided out so that each of these ratios should be unity.  In the full model there is reasonable agreement with this expectation in some cases.  However, it is clear that corrections to the symmetry are important for strange diquarks, as one would expect.  The worst disagreement in the full model occurs for the radially excited $1 ^3S_1\otimes 2S_\hlf$ multiplet.  The reason for this has been discussed earlier.
\begin{center}
\begin{table}[h]
\caption{Ratios of decay amplitudes involving the ground state strange diquark.  These are calculated in three different variations of the RP model: the full model, the spinless model, and the $\mu$SO model, which is the spinless RP model appended with an $\emph{ {\bf l}}_\lambda \cdot {\bf s}_l$ interaction.  The appropriate spin-counting factors have been divided out so that each of these ratios should be unity.
\label{XiSpinCount1+}}
\vspace{5mm}
\renewcommand{\arraystretch}{1.9}
\begin{tabular}{cc|ccc|ccc|ccc}

\hline \toprule
Multiplet & Decay Ratios & \multicolumn{3}{c|}{Full Model} & \multicolumn{3}{c|}{Spinless}& \multicolumn{3}{c}{$\mu$SO} \vspace{-3pt} \\ 
$n_d^{2s_d+1}(l_\rho)_{J_d}\otimes n_\lambda(l_\lambda)_{J_l}$ &${\cal M}(J_a,J_b,l)$ & $\Xi$&$\Xi_{c}$& $\Xi_{b}$& $\Xi$& $\Xi_{c}$& $\Xi_{b}$ & $\Xi$& $\Xi_{c}$& $\Xi_{b}$ 
 \\  \hline

\multirow{1}{*}{$1^{3}S_1\otimes 1P_\hlf$} &\large{$\frac{{\cal M}(\thlf,\thlf,0)}{{\cal M}(\hlf,\hlf,0)}$}  &
                               1.371 & 1.040 & 0.981 & 0.791 & 0.790 & 0.791 & 1.000 & 1.000 & 1.000 \\[+4pt] \hline

\multirow{4}{*}{$1 ^3S_1\otimes 1P_\thlf$} &\large{$\frac{1}{\sqrt{5}} \frac{{\cal M}(\thlf,\hlf,2)}{{\cal M}(\hlf,\thlf,2)}$}  &
                               0.763 & 0.630 & 0.621 & 0.791 & 0.791 & 0.791 & 1.000 & 1.000 & 1.000 \\[+4pt] 
                                 &\large{$\frac{2}{\sqrt{5}} \frac{{\cal M}(\thlf,\thlf,2)}{{\cal M}(\hlf,\thlf,2)}$} &
                               0.951 & 0.754 & 0.722 & 0.791 & 0.791 & 0.791 & 1.000 & 1.000 & 1.000 \\[+4pt] 
                                 &\large{$\sqrt{\frac{8}{15}} \frac{{\cal M}(\fhlf,\hlf,2)}{{\cal M}(\hlf,\thlf,2)}$} &   
                               0.910 & 0.829 & 0.854 & 1.061 & 1.061 & 1.061 & 1.000 & 1.000 & 1.000  \\[+4pt] 
                                 &\large{$\sqrt{\frac{7}{15}} \frac{{\cal M}(\fhlf,\thlf,2)}{{\cal M}(\hlf,\thlf,2)}$} &
                               1.240 & 0.999 & 0.999 & 1.061 & 1.061 & 1.061 & 1.000 & 1.000 & 1.000   \\[+4pt]  \hline

\multirow{3}{*}{$1 ^3S_1\otimes 2S_\hlf$} &\large{$\frac{1}{2\sqrt{2}} \frac{{\cal M}(\hlf,\thlf,1)}{{\cal M}(\hlf,\hlf,1)}$}  &
                               1.797 & 3.193 & 3.250 & 1.000 & 1.000 & 1.000 & 1.000 & 1.000 & 1.000 \\[+4pt] 
                                 &\large{$\frac{1}{2} \frac{{\cal M}(\thlf,\hlf,1)}{{\cal M}(\hlf,\hlf,1)}$} &
                               0.273 & 0.871 & 0.729 & 1.000 & 1.000 & 1.000 & 1.000 & 1.000 & 1.000 \\[+4pt] 
                                 &\large{$\frac{1}{\sqrt{5}} \frac{{\cal M}(\thlf,\thlf,1)}{{\cal M}(\hlf,\hlf,1)}$}&   
                               0.001 & 1.336 & 1.495 & 1.000 & 1.000 & 1.000 & 1.000 & 1.000 & 1.000  \\[+4pt]   \hline

\multirow{4}{*}{$1 ^3S_1\otimes 1D_\thlf$} &\large{$2\sqrt{2} \frac{{\cal M}(\hlf,\thlf,1)}{{\cal M}(\hlf,\hlf,1)}$}  &
                               1.305 & 1.374 & 1.381 & 1.000 & 1.000 & 1.000 & 1.000 & 1.000 & 1.000 \\[+4pt] 
                                 &\large{$2\sqrt{\frac{2}{5}} \frac{{\cal M}(\thlf,\hlf,1)}{{\cal M}(\hlf,\hlf,1)}$} &
                               1.125 & 1.022 & 0.993 & 0.894 & 0.894 & 0.894 & 1.000 & 1.000 & 1.000 \\[+4pt] 
                                 &\large{$\sqrt{2} \frac{{\cal M}(\thlf,\thlf,1)}{{\cal M}(\hlf,\hlf,1)}$} &   
                               1.515 & 1.404 & 1.381 & 0.894 & 0.894 & 0.894 & 1.000 & 1.000 & 1.000  \\[+4pt] 
                                 &\large{$\frac{2 \sqrt{2}}{3} \frac{{\cal M}(\fhlf,\thlf,1)}{{\cal M}(\hlf,\hlf,1)}$} &
                               1.270 & 1.383 & 1.269 & 0.725 & 0.723 & 0.719 & 0.992 & 0.979 & 0.973   \\[+4pt]  \hline

\multirow{4}{*}{$1 ^3S_1\otimes 1D_\fhlf$} &\large{$\sqrt{\frac{21}{5}} \frac{{\cal M}(\fhlf,\hlf,3)}{{\cal M}(\thlf,\thlf,3)}$}  &
                               0.532 & 0.675 & 0.657 & 0.824 & 0.831 & 0.835 & 1.009 & 1.021 & 1.028 \\[+4pt] 
                                 &\large{$\frac{\sqrt{21}}{4} \frac{{\cal M}(\fhlf,\thlf,3)}{{\cal M}(\thlf,\thlf,3)}$} &
                               0.744 & 0.926 & 0.868 & 0.824 & 0.831 & 0.835 & 1.009 & 1.021 & 1.028 \\[+4pt] 
                                 &\large{$\frac{\sqrt{7}}{2} \frac{{\cal M}(\shlf,\hlf,3)}{{\cal M}(\thlf,\thlf,3)}$} &   
                               0.628 & 0.724 & 0.765 & 1.118 & 1.118 & 1.118 & 1.000 & 1.000 & 1.000  \\[+4pt] 
                                 &\large{$\sqrt{\frac{7}{3}} \frac{{\cal M}(\fhlf,\thlf,2)}{{\cal M}(\hlf,\thlf,2)}$} &
                               1.017 & 0.995 & 1.005 & 1.118 & 1.118 & 1.118 & 1.000 & 1.000 & 1.000   \\[+4pt]
\hline
\toprule
\end{tabular}
\end{table}
\end{center}

As with the decays of DHBs, the results for spin-counting ratios in the spinless RP model show that spin-counting relations appear to be broken even when spin dependent forces are removed from the model.  This arises from the choice of the L--S basis as was discussed previously.  The inclusion of the $\rmb{l}_\lambda \cdot \rmb{s}_l$ interaction in the $\mu$SO model removes the spurious mixing which causes this disagreement.  The results for the spin-counting ratios in Table~\ref{XiSpinCount1+} in the $\mu$SO model agree with symmetry predictions to within a few percent.  This is a clear illustration that the primary mechanism in this model which breaks the heavy diquark spin symmetry is the mixing of multiplets, and this fact persists when strange quarks are treated as heavy quarks.

Tables \ref{XiNoPStripwidths} and \ref{XiNoPSsingwidths} show the pion-emission decay amplitudes obtained from the full RP model.  Tables \ref{XiMSOtripwidths} and \ref{XiMSOsingwidths} show analogous results for the $\mu$SO model.  In these tables the appropriate spin-counting factors have been divided out so that all of the amplitudes associated with a particular transition between HDS multiplets should be equal.  In the full RP model (Table~\ref{XiNoPStripwidths}), spin-counting relations for transitions from the $1^{1}P_1\otimes 1P_\frac{1}{2}$ multiplet to the $1^{1}P_1\otimes 1S_\frac{1}{2}$ multiplet agree reasonably well with the model, while the relations for transitions from the $1^{1}P_1\otimes 1P_\frac{3}{2}$ multiplet to the $1^{1}P_1\otimes 1S_\frac{1}{2}$ multiplet deviate more from the model predictions.  However, these relations improve as the mass of the strange diquark increases.  For flavor-antitriplet states in the full model (Table~\ref{XiNoPSsingwidths}), these relations perform more poorly, particularly for decays of the $1^{3}P_2\otimes 1P_\frac{3}{2}$ quadruplet to the $1^{3}P_2\otimes 1S_\frac{1}{2}$ doublet.  This disagreement is due to the significant amount of mixing to which these states are subject. Corrections to these predictions appear to be quite important for baryons with strange diquarks, particularly the $\Xi$. 
\begin{center}
\begin{table}[h]
\caption{HDS decay amplitudes of flavor-sextet strange states by pion-emission from the parent multiplet listed to the daughter multiplet listed.  The amplitudes are arranged in matrix form with the parent baryon's position in the multiplet corresponding to the column of the matrix and the daughter baryon's position in the multiplet corresponding to the row of the matrix.  These decay amplitudes are extracted from the $^3P_0$ model using quark model wave functions taken from the RP model \cite{RobertsPervin}.  Phase space has been divided off as well as the appropriate spin-counting factors.  Thus, all of the amplitudes, labeled by $A^l_{J_l, J^\prime_l}$, in each section of the table should be equal according to lowest order HDS predictions.  The entries in each set of parentheses are equal because of spin-counting arguments, the entries in different columns are equal because of the diquark flavor symmetry, and the entries involving different excitation states of the heavy diquark are equal because of the excitation symmetry.  The decay modes omitted are forbidden by spin-counting arguments.
\label{XiNoPStripwidths}}
\vspace{2mm}
\resizebox{7.07in}{!} {
\begin{tabular}{ccccccc}

\hline \toprule
\multirow{2}{*}{Parent Multiplet} & \multirow{2}{*}{$J_a^{P_a}$} & \multirow{2}{*}{Daughter Multiplet} & \multirow{2}{*}{$\qquad J_b^{P_b} \qquad$} & \multicolumn{3}{c}{HDS Amplitude $A^l_{J_l, J^\prime_l}$ (GeV$^{-l-\hlf}$)} \\ 
 & && & \large{$\Xi$} & \large{$\Xi_{c}$} & \large{$\Xi_{b}$} \\ \hline

 & & & & \multicolumn{3}{c}{\large{$A^0_{\hlf, \hlf}$}} \vspace{1mm} \\ \cline{5-7} 

$1^{3}S_1\otimes 1P_\frac{1}{2}$ & $(\hlf^-,\thlf^-)$           
& $1^{3}S_1\otimes 1S_\frac{1}{2}$ & $\left( \begin{array} {c} \hlf^+ \\ \thlf^+ \end{array} \right)$
& $\left( \begin{array} {c} 0.486 \;\;  \quad-\quad  \\  \quad-\quad  \;\; 0.667 \end{array} \right)$  
& $\left( \begin{array} {c} 0.628 \;\;  \quad-\quad  \\  \quad-\quad  \;\; 0.653 \end{array} \right)$  
& $\left( \begin{array} {c} 0.637 \;\;  \quad-\quad  \\  \quad-\quad  \;\; 0.625 \end{array} \right)$ \vspace{2mm} \\ 

$1^{1}P_1\otimes 1P_\frac{1}{2}$              & $(\hlf^+,\thlf^+)$           
& $1^{1}P_1\otimes 1S_\frac{1}{2}$ & $\left( \begin{array} {c} \hlf^- \\ \thlf^- \end{array} \right)$ 
& $\left( \begin{array} {c} 0.535 \;\;  \quad-\quad  \\  \quad-\quad  \;\; 0.635 \end{array} \right)$  
& $\left( \begin{array} {c} 0.564 \;\;  \quad-\quad  \\  \quad-\quad  \;\; 0.672 \end{array} \right)$  
& $\left( \begin{array} {c} 0.578 \;\;  \quad-\quad  \\  \quad-\quad  \;\; 0.678 \end{array} \right)$ \vspace{2mm} \\  

 & & & & \multicolumn{3}{c}{\large{$A^2_{\thlf, \hlf}$}} \vspace{1mm} \\ \cline{5-7} 

$1^{3}S_1\otimes 1P_\frac{3}{2}$              & $(\hlf^-,\thlf^-,\fhlf^-)$           
& $1^{3}S_1\otimes 1S_\frac{1}{2}$ & $\left( \begin{array} {c} \hlf^+ \\ \thlf^+ \end{array} \right)$ 
& $\left( \begin{array} {c} \quad-\quad   1.428 \;\; 1.703 \\ 1.871 \;\; 1.779 \;\; 2.321 \end{array} \right)$  
& $\left( \begin{array} {c} \quad-\quad   1.455 \;\; 1.912 \\ 2.308 \;\; 1.741 \;\; 2.305 \end{array} \right)$  
& $\left( \begin{array} {c} \quad-\quad   1.370 \;\; 1.884 \\ 2.205 \;\; 1.593 \;\; 2.202 \end{array} \right)$ \vspace{2mm} \\ 

$1^{1}P_1\otimes 1P_\frac{3}{2}$              & $(\hlf^+,\thlf^+,\fhlf^+)$           
& $1^{1}P_1\otimes 1S_\frac{1}{2}$ & $\left( \begin{array} {c} \hlf^- \\ \thlf^- \end{array} \right)$ 
& $\left( \begin{array} {c} \quad-\quad   1.679 \;\; 1.442 \\ 0.959 \;\; 1.971 \;\; 1.343 \end{array} \right)$  
& $\left( \begin{array} {c} \quad-\quad   1.939 \;\; 2.170 \\ 1.708 \;\; 1.962 \;\; 2.152 \end{array} \right)$  
& $\left( \begin{array} {c} \quad-\quad   1.852 \;\; 2.104 \\ 1.739 \;\; 1.875 \;\; 2.084 \end{array} \right)$ \vspace{2mm} \\  

 & & & & \multicolumn{3}{c}{\large{$A^1_{\hlf, \hlf}$}} \vspace{1mm} \\ \cline{5-7} 

$1^{3}S_1\otimes 2S_\frac{1}{2}$              & $(\hlf^+,\thlf^+)$           
& $1^{3}S_1\otimes 1S_\frac{1}{2}$ & $\left( \begin{array} {c} \hlf^+ \\ \thlf^+ \end{array} \right)$
& $\left( \begin{array} {c} 0.362 \;\;  0.099  \\  0.650  \;\; 0.000 \end{array} \right)$  
& $\left( \begin{array} {c} 0.088 \;\;  0.077  \\  0.281  \;\; 0.118 \end{array} \right)$  
& $\left( \begin{array} {c} 0.081 \;\;  0.059  \\  0.263  \;\; 0.121 \end{array} \right)$ \vspace{2mm} \\ 

 & & & & \multicolumn{3}{c}{\large{$A^1_{\thlf, \hlf}$}}  \vspace{1mm} \\ \cline{5-7}

$1^{3}S_1\otimes 1D_\frac{3}{2}$              & $(\hlf^+,\thlf^+,\fhlf^+)$           
& $1^{3}S_1\otimes 1S_\frac{1}{2}$ & $\left( \begin{array} {c} \hlf^+ \\ \thlf^+ \end{array} \right)$
& $\left( \begin{array} {c} 0.386 \;\; 0.434, \quad-\quad \\ 0.503 \;\; 0.584 \;\; 0.490 \end{array} \right)$  
& $\left( \begin{array} {c} 0.379 \;\; 0.387, \quad-\quad \\ 0.520 \;\; 0.532 \;\; 0.524 \end{array} \right)$  
& $\left( \begin{array} {c} 0.341 \;\; 0.339, \quad-\quad \\ 0.472 \;\; 0.472 \;\; 0.433 \end{array} \right)$ \vspace{2mm} \\ 

 & & & & \multicolumn{3}{c}{\large{$ A^3_{\fhlf, \hlf}$}}  \vspace{1mm} \\ \cline{5-7} 

$1^{3}S_1\otimes 1D_\frac{5}{2}$              & $(\thlf^+,\fhlf^+,\shlf^+)$           
& $1^{3}S_1\otimes 1S_\frac{1}{2}$ & $\left( \begin{array} {c} \hlf^+ \\ \thlf^+ \end{array} \right)$
& $\left( \begin{array} {c} \quad-\quad   0.782 \;\; 0.923 \\ 1.470 \;\; 1.094 \;\; 1.496 \end{array} \right)$  
& $\left( \begin{array} {c} \quad-\quad   0.882 \;\; 0.946 \\ 1.307 \;\; 1.210 \;\; 1.300 \end{array} \right)$  
& $\left( \begin{array} {c} \quad-\quad   0.745 \;\; 0.867 \\ 1.133 \;\; 0.983 \;\; 1.138 \end{array} \right)$ \vspace{2mm} \\ 
\hline
\toprule
\end{tabular}
}
\end{table}
\end{center}
\begin{center}
\begin{table}[h]
\caption{HDS decay amplitudes of flavor-antitriplet strange states by pion-emission from the parent multiplet listed to the daughter multiplet listed.  The amplitudes are arranged in matrix form with the parent baryon's position in the multiplet corresponding to the column of the matrix and the daughter baryon's position in the multiplet corresponding to the row of the matrix.  These decay amplitudes are extracted from the $^3P_0$ model using quark model wave functions taken from the RP model \cite{RobertsPervin}.  Phase space has been divided off as well as the appropriate spin-counting factors.  Thus, all of the amplitudes, labeled by $A^l_{J_l, J^\prime_l}$, in each section of the table should be equal according to lowest order HDS predictions.  The entries in each set of parentheses are equal because of spin-counting arguments, and the entries involving different excitation states of the heavy diquark are equal because of the excitation symmetry.  These entries should also equal the corresponding amplitudes for flavor-triplet states.  The decay modes omitted are forbidden by spin-counting arguments.
\label{XiNoPSsingwidths}}
\vspace{2mm}
\resizebox{7.07in}{!} {
\begin{tabular}{c c c c c c}

\hline \toprule
\multirow{2}{*}{Parent Multiplet} & \multirow{2}{*}{$J_a^{P_a}$} & \multirow{2}{*}{Daughter Multiplet} & \multirow{2}{*}{$\qquad J_b^{P_b} \qquad$} & \multicolumn{2}{c}{HDS Amplitude $A^l_{J_l, J^\prime_l}$ (GeV$^{-l-\hlf}$)} \\ 
 & && &  \large{$\Xi^\prime_{c}$}&  \large{$\Xi^\prime_{b}$}  \\ \hline
 & & & & \multicolumn{2}{c}{\large{$A^0_{\hlf, \hlf}$}} \vspace{1mm} \\ \cline{5-6} 

$1^{1}S_0\otimes 1P_\frac{1}{2}$              & $(\hlf^-)$           
& $1^{1}S_0\otimes 1S_\frac{1}{2}$ & $(\hlf^+)$ & (0.772) & (0.778) \vspace{1mm} \\ 

$1^{3}P_0\otimes 1P_\frac{1}{2}$              & $(\hlf^+)$           
& $1^{3}P_0\otimes 1S_\frac{1}{2}$ & $(\hlf^-)$ & (0.701) & (0.688) \vspace{1mm} \\ 

$1^{3}P_1\otimes 1P_\frac{1}{2}$              & $(\hlf^+,\thlf^+)$           
& $1^{3}P_1\otimes 1S_\frac{1}{2}$ & $\left( \begin{array} {c} \hlf^- \\ \thlf^- \end{array} \right)$
& $\left( \begin{array} {c} 0.553 \;\;  \quad-\quad  \\  \quad-\quad  \;\; 0.525 \end{array} \right)$ 
& $\left( \begin{array} {c} 0.615 \;\;  \quad-\quad  \\  \quad-\quad  \;\; 0.480 \end{array} \right)$ \vspace{2mm} \\ 

$1^{3}P_2\otimes 1P_\frac{1}{2}$              & $(\thlf^+,\fhlf^+)$           
& $1^{3}P_2\otimes 1S_\frac{1}{2}$ & $\left( \begin{array} {c} \thlf^- \\ \fhlf^- \end{array} \right)$
& $\left( \begin{array} {c} 0.495 \;\;  \quad-\quad  \\  \quad-\quad  \;\; 0.620 \end{array} \right)$
& $\left( \begin{array} {c} 0.479 \;\;  \quad-\quad  \\  \quad-\quad  \;\; 0.613 \end{array} \right)$ \vspace{2mm} \\ 

 & & & & \multicolumn{2}{c}{\large{$A^2_{\thlf, \hlf}$}}  \vspace{1mm} \\ \cline{5-6}

$1^{1}S_0\otimes 1P_\frac{3}{2}$              & $(\thlf^-)$           
& $1^{1}S_0\otimes 1S_\frac{1}{2}$ & $(\hlf^+)$ & (2.126) & (2.076) \vspace{1mm} \\ 

$1^{3}P_0\otimes 1P_\frac{3}{2}$              & $(\thlf^+)$           
& $1^{3}P_0\otimes 1S_\frac{1}{2}$ & $(\hlf^-)$ & (1.036) & (1.031) \vspace{1mm} \\ 

$1^{3}P_1\otimes 1P_\frac{3}{2}$              & $(\hlf^+,\thlf^+,\fhlf^+)$           
& $1^{3}P_1\otimes 1S_\frac{1}{2}$ & $\left( \begin{array} {c} \hlf^- \\ \thlf^- \end{array} \right)$ 
& $\left( \begin{array} {c} \quad-\quad   1.027 \;\; 1.149 \\ 1.851 \;\; 1.391 \;\; 1.116 \end{array} \right)$
& $\left( \begin{array} {c} \quad-\quad   1.832 \;\; 1.023 \\ 1.769 \;\; 1.439 \;\; 1.640 \end{array} \right)$ \vspace{2mm} \\  

$1^{3}P_2\otimes 1P_\frac{3}{2}$              & $(\hlf^+,\thlf^+,\fhlf^+,\shlf^+)$           
& $1^{3}P_2\otimes 1S_\frac{1}{2}$ & $\left( \begin{array} {c} \thlf^- \\ \fhlf^- \end{array} \right)$ 
& $\left( \begin{array} {c} 4.058 \;\; 1.163 \;\; 0.386 \;\; 1.866 \\ 
                            1.726 \;\; 1.729 \;\; 1.118 \;\; 2.355 \end{array} \right)$
& $\left( \begin{array} {c} 3.555 \;\; 1.108 \;\; 1.856 \;\; 1.855 \\ 
                            1.758 \;\; 1.696 \;\; 1.578 \;\; 2.254 \end{array} \right)$  \vspace{2mm} \\  

 & & & & \multicolumn{2}{c}{\large{$A^1_{\hlf, \hlf}$}} \vspace{1mm} \\ \cline{5-6}

$1^{1}S_0\otimes 2S_\frac{1}{2}$              & $(\hlf^+)$           
& $1^{1}S_0\otimes 1S_\frac{1}{2}$ & $(\hlf^+)$ & (0.090)  & (0.108) \vspace{1mm} \\ 

 & & & & \multicolumn{2}{c}{\large{$A^1_{\thlf, \hlf}$}} \vspace{1mm} \\ \cline{5-6}

$1^{1}S_0\otimes 1D_\frac{3}{2}$              & $(\thlf^+)$           
&$1^{1}S_0\otimes 1S_\frac{1}{2}$ & $(\hlf^+)$ & (0.462) & (0.422)  \vspace{1mm} \\ 

 & & & & \multicolumn{2}{c}{\large{$A^3_{\fhlf, \hlf}$}} \vspace{1mm} \\ \cline{5-6}

$1^{1}S_0\otimes 1D_\frac{5}{2}$              & $(\fhlf^+)$           
& $1^{1}S_0\otimes 1S_\frac{1}{2}$ & $(\hlf^+)$ & (1.132) & (1.042)  \vspace{1mm} \\ 
\hline
\toprule
\end{tabular}}
\end{table}
\end{center}
\begin{center}
\begin{table}[h]
\caption{Same as Table \ref{XiNoPStripwidths} but using wave functions from the spinless RP model including the $\emph{ {\bf l}}_\lambda \cdot {\bf s}_l$ interaction.
\label{XiMSOtripwidths}}
\vspace{2mm}
\resizebox{7.07in}{!} {
\begin{tabular}{c c c c c c c}

\hline \toprule
\multirow{2}{*}{Parent Multiplet} & \multirow{2}{*}{$J_a^{P_a}$} & \multirow{2}{*}{Daughter Multiplet} & \multirow{2}{*}{$\qquad J_b^{P_b} \qquad$} & \multicolumn{3}{c}{HDS Amplitude $A^l_{J_l, J^\prime_l}$ (GeV$^{-l-\hlf}$)} \\ 
 & && & \large{$\Xi$} & \large{$\Xi_{c}$} & \large{$\Xi_{b}$} \\ \hline

 & & & & \multicolumn{3}{c}{\large{$A^0_{\hlf, \hlf}$}} \vspace{1mm} \\ \cline{5-7} 

$1^{3}S_1\otimes 1P_\frac{1}{2}$ & $(\hlf^-,\thlf^-)$           
& $1^{3}S_1\otimes 1S_\frac{1}{2}$ & $\left( \begin{array} {c} \hlf^+ \\ \thlf^+ \end{array} \right)$
& $\left( \begin{array} {c} 0.814 \;\;  \quad-\quad  \\  \quad-\quad  \;\; 0.814 \end{array} \right)$  
& $\left( \begin{array} {c} 0.846 \;\;  \quad-\quad  \\  \quad-\quad  \;\; 0.846 \end{array} \right)$  
& $\left( \begin{array} {c} 0.847 \;\;  \quad-\quad  \\  \quad-\quad  \;\; 0.847 \end{array} \right)$ \vspace{2mm} \\   

$1^{1}P_1\otimes 1P_\frac{1}{2}$              & $(\hlf^+,\thlf^+)$           
& $1^{1}P_1\otimes 1S_\frac{1}{2}$ & $\left( \begin{array} {c} \hlf^- \\ \thlf^- \end{array} \right)$ 
& $\left( \begin{array} {c} 0.623 \;\;  \quad-\quad  \\  \quad-\quad  \;\; 0.730 \end{array} \right)$  
& $\left( \begin{array} {c} 0.636 \;\;  \quad-\quad  \\  \quad-\quad  \;\; 0.748 \end{array} \right)$  
& $\left( \begin{array} {c} 0.636 \;\;  \quad-\quad  \\  \quad-\quad  \;\; 0.751 \end{array} \right)$ \vspace{2mm} \\   

 & & & & \multicolumn{3}{c}{\large{$A^2_{\thlf, \hlf}$}} \vspace{1mm} \\ \cline{5-7} 

$1^{3}S_1\otimes 1P_\frac{3}{2}$              & $(\hlf^-,\thlf^-,\fhlf^-)$           
& $1^{3}S_1\otimes 1S_\frac{1}{2}$ & $\left( \begin{array} {c} \hlf^+ \\ \thlf^+ \end{array} \right)$ 
& $\left( \begin{array} {c} \quad-\quad   2.303 \;\; 2.303 \\ 2.303 \;\; 2.303 \;\; 2.303 \end{array} \right)$  
& $\left( \begin{array} {c} \quad-\quad   2.178 \;\; 2.178 \\ 2.178 \;\; 2.178 \;\; 2.178 \end{array} \right)$  
& $\left( \begin{array} {c} \quad-\quad   2.102 \;\; 2.102 \\ 2.102 \;\; 2.102 \;\; 2.102 \end{array} \right)$ \vspace{2mm} \\   

$1^{1}P_1\otimes 1P_\frac{3}{2}$              & $(\hlf^+,\thlf^+,\fhlf^+)$           
& $1^{1}P_1\otimes 1S_\frac{1}{2}$ & $\left( \begin{array} {c} \hlf^- \\ \thlf^- \end{array} \right)$ 
& $\left( \begin{array} {c} \quad-\quad   2.110 \;\; 2.392 \\ 1.978 \;\; 2.110 \;\; 2.392 \end{array} \right)$  
& $\left( \begin{array} {c} \quad-\quad   1.975 \;\; 2.266 \\ 1.860 \;\; 1.975 \;\; 2.266 \end{array} \right)$  
& $\left( \begin{array} {c} \quad-\quad   1.891 \;\; 2.193 \\ 1.792 \;\; 1.891 \;\; 2.193 \end{array} \right)$ \vspace{2mm} \\   

 & & & & \multicolumn{3}{c}{\large{$A^1_{\hlf, \hlf}$}} \vspace{1mm} \\ \cline{5-7} 

$1^{3}S_1\otimes 2S_\frac{1}{2}$              & $(\hlf^+,\thlf^+)$           
& $1^{3}S_1\otimes 1S_\frac{1}{2}$ & $\left( \begin{array} {c} \hlf^+ \\ \thlf^+ \end{array} \right)$
& $\left( \begin{array} {c} 0.177 \;\;  0.177  \\  0.177  \;\; 0.177 \end{array} \right)$  
& $\left( \begin{array} {c} 0.156 \;\;  0.156  \\  0.156  \;\; 0.156 \end{array} \right)$  
& $\left( \begin{array} {c} 0.127 \;\;  0.127  \\  0.127  \;\; 0.127 \end{array} \right)$ \vspace{2mm} \\   

 & & & & \multicolumn{3}{c}{\large{$A^1_{\thlf, \hlf}$}} \vspace{1mm} \\ \cline{5-7} 

$1^{3}S_1\otimes 1D_\frac{3}{2}$              & $(\hlf^+,\thlf^+,\fhlf^+)$           
& $1^{3}S_1\otimes 1S_\frac{1}{2}$ & $\left( \begin{array} {c} \hlf^+ \\ \thlf^+ \end{array} \right)$
& $\left( \begin{array} {c} 0.555 \;\; 0.555, \quad-\quad \\ 0.555 \;\; 0.555 \;\; 0.551 \end{array} \right)$  
& $\left( \begin{array} {c} 0.565 \;\; 0.565, \quad-\quad \\ 0.565 \;\; 0.565 \;\; 0.553 \end{array} \right)$  
& $\left( \begin{array} {c} 0.525 \;\; 0.525, \quad-\quad \\ 0.525 \;\; 0.525 \;\; 0.511 \end{array} \right)$ \vspace{2mm} \\

 & & & & \multicolumn{3}{c}{\large{$A^3_{\fhlf, \hlf}$}} \vspace{1mm} \\ \cline{5-7}  

$1^{3}S_1\otimes 1D_\frac{5}{2}$              & $(\thlf^+,\fhlf^+,\shlf^+)$           
& $1^{3}S_1\otimes 1S_\frac{1}{2}$ & $\left( \begin{array} {c} \hlf^+ \\ \thlf^+ \end{array} \right)$
& $\left( \begin{array} {c} \quad-\quad   1.311 \;\; 1.300 \\ 1.300 \;\; 1.311 \;\; 1.300 \end{array} \right)$  
& $\left( \begin{array} {c} \quad-\quad   1.145 \;\; 1.122 \\ 1.122 \;\; 1.145 \;\; 1.122 \end{array} \right)$  
& $\left( \begin{array} {c} \quad-\quad   1.027 \;\; 0.999 \\ 0.999 \;\; 1.027 \;\; 0.999 \end{array} \right)$ \vspace{2mm} \\   
\hline
\toprule
\end{tabular}
}
\end{table}
\end{center}
\begin{center}
\begin{table}[h]
\caption{Same as Table \ref{XiNoPSsingwidths} but using wave functions from the spinless RP model including the $\emph{ {\bf l}}_\lambda \cdot {\bf s}_l$ interaction.
\label{XiMSOsingwidths}}
\vspace{2mm}
\resizebox{7.07in}{!} {
\begin{tabular}{c c c c c c}

\hline \toprule
\multirow{2}{*}{Parent Multiplet} & \multirow{2}{*}{$J_a^{P_a}$} & \multirow{2}{*}{Daughter Multiplet} & \multirow{2}{*}{$\qquad J_b^{P_b} \qquad$} & \multicolumn{2}{c}{HDS Amplitude $A^l_{J_l, J^\prime_l}$ (GeV$^{-l-\hlf}$)} \\ 
 & && &  \large{$\Xi^\prime_{c}$}&  \large{$\Xi^\prime_{b}$}  \\ \hline

 & & & & \multicolumn{2}{c}{\large{$A^0_{\hlf, \hlf}$}} \vspace{1mm} \\ \cline{5-6} 

$1^{1}S_0\otimes 1P_\frac{1}{2}$              & $(\hlf^-)$           
& $1^{1}S_0\otimes 1S_\frac{1}{2}$ & $(\hlf^+)$ & (0.850) & (0.845) \vspace{1mm} \\ 

$1^{3}P_0\otimes 1P_\frac{1}{2}$              & $(\hlf^+)$           
& $1^{3}P_0\otimes 1S_\frac{1}{2}$ & $(\hlf^-)$ & (0.254) & (0.256) \vspace{1mm} \\ 

$1^{3}P_1\otimes 1P_\frac{1}{2}$              & $(\hlf^+,\thlf^+)$           
& $1^{3}P_1\otimes 1S_\frac{1}{2}$ & $\left( \begin{array} {c} \hlf^- \\ \thlf^- \end{array} \right)$
& $\left( \begin{array} {c} 0.216 \;\;  \quad-\quad  \\  \quad-\quad  \;\; 0.500 \end{array} \right)$ 
& $\left( \begin{array} {c} 0.217 \;\;  \quad-\quad  \\  \quad-\quad  \;\; 0.502 \end{array} \right)$ \vspace{2mm} \\ 

$1^{3}P_2\otimes 1P_\frac{1}{2}$              & $(\thlf^+,\fhlf^+)$           
& $1^{3}P_2\otimes 1S_\frac{1}{2}$ & $\left( \begin{array} {c} \thlf^- \\ \fhlf^- \end{array} \right)$
& $\left( \begin{array} {c} 0.425 \;\;  \quad-\quad  \\  \quad-\quad  \;\; 0.781 \end{array} \right)$
& $\left( \begin{array} {c} 0.425 \;\;  \quad-\quad  \\  \quad-\quad  \;\; 0.785 \end{array} \right)$ \vspace{2mm} \\   

 & & & & \multicolumn{2}{c}{\large{$A^2_{\thlf, \hlf}$}}  \vspace{1mm} \\ \cline{5-6}

$1^{1}S_0\otimes 1P_\frac{3}{2}$              & $(\thlf^-)$           
& $1^{1}S_0\otimes 1S_\frac{1}{2}$ & $(\hlf^+)$ & (2.185) & (2.110) \vspace{1mm} \\ 

$1^{3}P_0\otimes 1P_\frac{3}{2}$              & $(\thlf^+)$           
& $1^{3}P_0\otimes 1S_\frac{1}{2}$ & $(\hlf^-)$ & (1.673) & (1.616) \vspace{1mm} \\ 

$1^{3}P_1\otimes 1P_\frac{3}{2}$              & $(\hlf^+,\thlf^+,\fhlf^+)$           
& $1^{3}P_1\otimes 1S_\frac{1}{2}$ & $\left( \begin{array} {c} \hlf^- \\ \thlf^- \end{array} \right)$ 
& $\left( \begin{array} {c} \quad-\quad   1.992 \;\; 1.541 \\ 1.756 \;\; 0.895 \;\; 2.176 \end{array} \right)$
& $\left( \begin{array} {c} \quad-\quad   1.934 \;\; 1.493 \\ 1.697 \;\; 0.868 \;\; 2.106 \end{array} \right)$ \vspace{2mm} \\ 

$1^{3}P_2\otimes 1P_\frac{3}{2}$              & $(\hlf^+,\thlf^+,\fhlf^+,\shlf^+)$           
& $1^{3}P_2\otimes 1S_\frac{1}{2}$ & $\left( \begin{array} {c} \thlf^- \\ \fhlf^- \end{array} \right)$ 
& $\left( \begin{array} {c} 3.185 \;\; 0.672 \;\; 2.077 \;\; 2.069 \\ 
                            1.730 \;\; 1.462 \;\; 1.693 \;\; 2.247 \end{array} \right)$
& $\left( \begin{array} {c} 3.099 \;\; 0.647 \;\; 2.008 \;\; 2.002 \\ 
                            1.687 \;\; 1.414 \;\; 1.641 \;\; 2.180 \end{array} \right)$  \vspace{2mm} \\  

 & & & & \multicolumn{2}{c}{\large{$A^1_{\hlf, \hlf}$}} \vspace{1mm} \\ \cline{5-6}

$1^{1}S_0\otimes 2S_\frac{1}{2}$              & $(\hlf^+)$           
& $1^{1}S_0\otimes 1S_\frac{1}{2}$ & $(\hlf^+)$ & (0.156)  & (0.115) \vspace{1mm} \\ 

 & & & & \multicolumn{2}{c}{\large{$A^1_{\thlf, \hlf}$}} \vspace{1mm} \\ \cline{5-6}

$1^{1}S_0\otimes 1D_\frac{3}{2}$              & $(\thlf^+)$           
&$1^{1}S_0\otimes 1S_\frac{1}{2}$ & $(\hlf^+)$ & (0.563) & (0.515)  \vspace{1mm} \\ 

 & & & & \multicolumn{2}{c}{\large{$A^3_{\fhlf, \hlf}$}} \vspace{1mm} \\ \cline{5-6}

$1^{1}S_0\otimes 1D_\frac{5}{2}$              & $(\fhlf^+)$           
& $1^{1}S_0\otimes 1S_\frac{1}{2}$ & $(\hlf^+)$ & (1.152) & (1.035)  \vspace{1mm} \\ 
\hline
\toprule
\end{tabular}}
\end{table}
\end{center}

The HDS amplitudes (see Eq.~(\ref{HDS})) obtained using the $\mu$SO model are shown in Tables \ref{XiMSOtripwidths} and \ref{XiMSOsingwidths}.  For flavor-sextet decays involving an excited diquark, spin-counting relations work to within 30\% (Table \ref{XiMSOtripwidths}).  Departures from HDS expectations are due to mixing induced by pairwise confining forces.  The impact of these forces is even more conspicuous for flavor-antitriplet decays involving a P-wave heavy diquark (Table \ref{XiMSOsingwidths}).  Here, the worst disagreement occurs for decays of the $1^{3}P_2\otimes 1P_\frac{3}{2}$ quadruplet to the $1^{3}P_2\otimes 1S_\frac{1}{2}$ doublet, where the HDS amplitude for the decay of the $\hlf^+$ state to the $\thlf^-$ state differs from HDS amplitude for the decay of the $\thlf^+$ state to the $\thlf^-$ state by a factor of 4.74 for the $\Xi^\prime_c$ and a factor of 4.79 for the $\Xi^\prime_b$.  These results are quite similar to those obtained in the case of DHBs, Tables \ref{NoPStripwidths} to 
\ref{MSOsingwidths}

\subsubsection{Excitation Symmetry}

The amplitudes in Table~\ref{XiNoPStripwidths} and Table~\ref{XiNoPSsingwidths} are expected to be independent of the excitation state of the heavy diquark.  Thus, all of the amplitudes corresponding to the same transition in the light degrees of freedom should be equal.  For instance, the amplitudes in the first two rows of Table~\ref{XiNoPStripwidths} should be equal.  In the same way, the amplitudes in rows three and four should be equal.  In Table~\ref{XiNoPSsingwidths} the amplitudes in the first four rows should be equal, and the amplitudes in rows five to eight should be equal.  

For the $\Xi$ baryon (Table~\ref{XiNoPStripwidths}) there are significant departures from excitation symmetry expectations.  However, the results improve for the $\Xi_c$ and $\Xi_b$ baryons.  Therefore, this discrepancy illustrates the importance of spin dependent corrections for baryons with strange diquarks.  For the flavor anti-triplet states (Table~\ref{XiNoPSsingwidths}), the effects of this mixing are more pronounced.  The reason for this was discussed in detail in the discussion of DHB results.  

The excitation symmetry and the decoupling of the total angular momentum of the strange diquark suggest that the strong decay amplitudes will also be independent of the statistics of the strange diquark.  Therefore,  the HDS amplitudes in Table~\ref{XiNoPSsingwidths} should be equal to  the corresponding amplitudes in the $\Xi_c$ and $\Xi_b$ columns of Table~\ref{XiNoPStripwidths}.    In some cases there is reasonable agreement between these amplitudes.  However, corrections due to mixing seem to be very important.  There are also instances where these mixing effects appear to be smaller for $\Xi_c$/$\Xi^\prime_c$ amplitudes than for $\Xi_b$/$\Xi^\prime_b$ amplitudes.  In reference \cite{RobertsPervin} it was pointed out that hyperfine mixing decreases when the difference in the masses of the quarks comprising the diquark decreases. This is an example of that feature of quark models.

Tables~\ref{XiMSOtripwidths} and \ref{XiMSOsingwidths} show the results obtained using the $\mu SO$ model.  The departures from the diquark excitation symmetry are comparable to the case of heavy diquarks, Tables \ref{MSOtripwidths} and \ref{MSOsingwidths}.  In the cases where there is good agreement with symmetry expectations pairwise mixing is small.  The symmetry between flavor-antitriplet and flavor-sextet states can be seen by comparing the HDS amplitudes in Table~\ref{XiMSOsingwidths} to the corresponding amplitudes in the $\Xi_c$ and $\Xi_b$ columns of Table~\ref{XiMSOtripwidths}.

\subsubsection{Diquark Flavor Symmetry} \label{StrangeFlavor}

It is expected that as the mass of the strange diquark becomes very large the decay amplitudes will not depend on the flavor of the diquark.  As was discussed previously for the case of heavy diquarks, the amplitudes tend to decrease as the mass of the diquark increases.  This is true for most instances in both the full RP model (Table~\ref{XiNoPStripwidths} and Table~\ref{XiNoPSsingwidths}) and the $\mu SO$ model (Table~\ref{XiMSOtripwidths} and Table~\ref{XiMSOsingwidths}).  The $\mu SO$ model results are in much better agreement with this expectation than the results from the full model.  These results are consistent with those for DHBs, Tables \ref{NoPStripwidths} to \ref{MSOsingwidths}.

The precision to which the diquark flavor symmetry emerges for strange diquarks is surprising because the mass of the strange quark is not large enough for one to rigorously treat it as a heavy quark ($m_s \sim \Lambda_{QCD}$).  A comparison of the results from the full RP model to the results from the $\mu$SO model seems to indicate that spin interactions are the primary reason that this symmetry is broken for transitions of states containing an excited diquark.  Therefore, a comprehensive application of this symmetry to strange baryons may require some spin-dependent corrections.  However, it may be feasible to treat these corrections in a systematic but phenomenological way.  This raises the possibility of comparing decay amplitudes of the $\Xi$ (Table~\ref{XiMSOtripwidths}) to decay amplitudes of the $\Xi_{bb}$ (Table~\ref{MSOtripwidths}) in the $\mu SO$ model.  A few of these amplitudes are nearly equal, but in most cases there are appreciable deviations.  One should keep in mind, however, that dividing off phase space from the $^3P_0$ model amplitudes does not remove all of the heavy baryon mass dependence (see section \ref{PhaseSpace}). 

\section{Conclusions}
We have derived the consequences of HDS for strong decays.  The fundamental principle behind this symmetry is that, in the heavy quark limit, the two heavy quarks inside of a DHB form a heavy diquark in an antitriplet color configuration whose total angular momentum and flavor decouple from the light degrees of freedom.  This causes the DHB spectrum to consist of degenerate multiplets similar to those of HQS.  Strong decays of DHBs by light meson emission are subject to a selection rule which forbids transitions that alter the quantum numbers of the heavy diquark.  The allowed transitions between multiplets are subject to spin-counting rules that relate ratios of partial widths.  The amplitudes for these transitions are also independent of the excitation state of the heavy diquark which is a spectator in the process.  The decoupling of the flavor of the heavy diquark provides a simple way to relate the strong decays of $\Xi_{cc}$, $\Xi_{bc}$, and $\Xi_{bb}$ baryons.  The fact that the color structure of the heavy diquark is the same as a heavy antiquark implies the existence of a superflavor symmetry that relates the strong decay amplitudes of singly heavy mesons to those of DHBs.  HDS symmetry is an extension of the superflavor symmetry of Savage and Wise \cite{SavageSpectrum,eakinsroberts} in which the spin and flavor of the heavy diquark decouples from the light degrees of freedom but the heavy diquark cannot be excited.

We have calculated the pion-emission strong decay widths of DHBs using the $^3P_0$ model for transitions and wave functions from the RP model.  The RP model was not explicitly constructed with these symmetries and contains pairwise confining forces and spin dependent interactions which violate the symmetries.  Strong decays of DHBs are sensitive to the mixing introduced by these effects.  Even when the spin dependent interactions are removed from the model, the symmetry is still violated by pairwise interactions.  Some of this is due to an arbitrary choice of basis, but including a small $l_\lambda \cdot s_l$ interaction in the spinless RP model (the $\mu$SO model) removes most of these effects, providing a more accurate representation of the infinitely heavy quark limit.  However, states which consist of both an orbital excitation of the heavy diquark and an orbital excitation of the light degrees of freedom are still subject to mixing due to the existence of pairwise confining forces in the model.  This mixing does not diminish in the heavy quark limit, and violates HDS.  This characteristic is not unique to the RP model, but rather, {\it any} nonrelativistic model that implements pairwise confining forces will exhibit this pathology.  The $^3P_0$ model treats the heavy diquark as a spectator in the strong decay process, and therefore, explicitly respects HDS.  Moreover, we have argued that any strong decay model that implements a spectator assumption will respect HDS.  

Our comparison of the RP model and $^3P_0$ model to HDS constraints showed reasonable agreement for DHBs.  The results for decay widths of DHB resonances by pion-emission to the ground state multiplet show a clear suppression for parent states containing an excited heavy diquark.  This is a result of the heavy diquark selection rule.  Many of these suppressed widths are less than a KeV, implying that electromagnetic transitions could be very important for resonances consisting of a heavy diquark excitation.  Some of the allowed decays are rather large in the model.  These are all S-wave transitions, and this may be an indication that the $^3P_0$ model does not appropriately capture the physics of S-wave decays.  Another possibility is that the $^3P_0$ constant has not been adjusted to an appropriate value.  If this is the case, then the suppressed widths that we have calculated should be even smaller.  The spin-counting relations for DHBs agree with model predictions to about 10\%, typically, but some of the spin-counting relations are in worse agreement illustrating the need for $1/M_Q$ corrections.  The results for spin-counting ratios in the $\mu$SO model agree quite well, indicating that the most important corrections are due to spin interactions.  The heavy diquark excitation symmetry emerges in the full RP model at an accuracy of 30\% or better for flavor-triplet DHBs.  For flavor singlet DHBs, the agreement is worse.  However, mixing from spin interactions is larger for these states.  A systematic treatment of these corrections to HDS seems to be necessary for this symmetry to be useful.  HDS flavor symmetry constraints agree quite well with RP model predictions.  These relations are even more precise in the $\mu$SO model.  This is another indication of an emergent quark-heavy diquark structure for DHBs in the RP model. 

We compared the results of the model for $\Xi$, $\Xi_c$, and $\Xi_b$ baryons to HDS by treating the strange quark as a heavy quark.  The suppression of pion decays that are forbidden by the heavy diquark selection rule is not as pronounced as it is for DHBs; however, experimental results for the lowest negative parity $\Xi$ resonances, which consist of diquark excitations in this treatment, indicate that the model predicts pion transitions that are much too large for these states.  The results for spin-counting ratios in the full RP model involving strange diquarks show that spin dependent corrections to HDS are very important for these baryons, as one would expect.  On the other hand, the spin-counting ratios from the $\mu$SO model agree precisely with lowest order HDS predictions.  This seems to indicate that the appreciable size of the strange diquark does not have a significant impact on these relations.  The diquark excitation symmetry for strange diquarks is in worse agreement with lowest order HDS than in the case of DHBs.  In the $\mu$SO model, the $SU(3)_{heavy}$ flavor-sextet results disagree with this HDS constraint by 33\% or less, but the flavor-antitriplet states violate this constraint by as much as 400\%.  These discrepancies are due to mixing from pairwise confining forces in the model.  There may be significant corrections due to the appreciable size of the strange diquark for these symmetry predictions, but those effects cannot be estimated by this model because of the pronounced mixing due to pairwise confining forces.  The model predictions agree with the diquark flavor symmetry quite well.  Again, these results are very similar to those for DHBs.  

We conclude by commenting on the scope of the Heavy Diquark Symmetry, if it can indeed be applied to strange diquarks.  In the RP model there are a large number of states calculated for the $\Xi$, $\Xi_c$, $\Xi_b$, $\Xi_{cc}$, $\Xi_{bc}$, $\Xi_{bb}$ flavors of baryons.  There are a multitude of partial widths pertaining to the pion-emission strong decays of these states.  HDS relates 240 partial widths to just five amplitudes, and it explains the suppression of the remaining pion decay channels. The tension between the application of HDS to $\Xi_c$/$\Xi_b$ cascades and the application of HQS may prove to be quite interesting.  On the one hand, if one treats the strange quark as a `heavy' quark, it is not heavy enough to significantly suppress interactions involving its spin.  However, it might be possible to (phenomenologically) correct for these effects in a systematic way provided that these particles have a prominent $sc$ or $sb$  diquark structure.  On the other hand, if one treats the strange quark as a light quark and applies HQS, the mass of the strange quark cannot be neglected. The application of these ideas to $\Xi$ resonances may provide new insights about their spectroscopy not accessible otherwise.  We speculate that the symmetries discussed here could also provide insight into the electromagnetic decays of DHBs.  Such a treatment could be particularly useful if the pion decays of the low lying $\Xi_{cc}$, $\Xi_{bc}$, and $\Xi_{bb}$ resonances are indeed suppressed as much as this calculation indicates.  This is a topic for future work.

\section*{Acknowledgment} We gratefully acknowledge the support of the Department of Physics, the College of Arts and Sciences, and the Office of Research at Florida State University.
This research is supported by the U.S. Department of Energy under contract
DE-SC0002615.

\appendix
\section{Derivation of Spin-Counting Factor from $^3P_0$ Amplitudes}
The $^3P_0$ amplitude may be written in a form that is valid for both heavy mesons and doubly heavy baryons \cite{Roberts-Brac} 
\begin{eqnarray} \label{3P0amp}
\nonumber {\cal M}&=&\frac{1}{3\sqrt{3}} (-1)^{l+l_a} {\cal F} {\cal R} \sum_{j_{bc}} (-1)^{j_a-j_{bc}} \left[ \begin{array} {ccc} J_d & \hlf   & j_b \\ 
                            s_q & \hlf & S_c \\
                            j_a & 1  & j_{bc} \end{array} \right] \ \\
\nonumber         &\times& 
 \sum_{l_{bc}} (-1)^{l_{bc}} \left[ \begin{array} {ccc} j_b & l_b   & J_b \\ 
                            S_c & l_c & J_c \\
                            j_{bc} & l_{bc}  & J_{bc} \end{array} \right] 
                            \sum_{L} \hat{L}^2 \left\{ \begin{array} {ccc} j_a & l_a   & J_a \\ 
                            L & j_{bc}  & 1 \end{array} \right\} \ \\
        &\times&  
                           \left\{ \begin{array} {ccc} j_{bc} & l_{bc}   & J_{bc} \\ 
                            l & J_a  & L \end{array} \right\} \epsilon(l_b,l_c,l_{bc},l,l_a,L)     \ \  
\end{eqnarray}
where ${\cal F}$ is the flavor overlap, ${\cal R}$ is the overlap of the spectator part of the wave function, $\epsilon(l_b,l_c,l_{bc},l,l_a,L)$ is the spatial part of the integral over the light degrees of freedom, $s_q=\hlf$ is the spin of the light quark which participates in the decay process, $S_c=s_{1c}+s_{2c}$ is the spin of the light daughter meson, and $J_d$ is the total angular momentum of the heavy degrees of freedom.  The `square bracket' notation for the 9--j symbol is defined as
\beq
\left[ \begin{array} {ccc} a & b & c \\ 
                           d & e & f \\
                           g & h & i \end{array} \right] =
\hat{c}\hat{f}\hat{g}\hat{h}\hat{i} 
\left\{ \begin{array} {ccc} a & b & c \\ 
                           d & e & f \\
                           g & h & i \end{array} \right\},
\eeq
with $\hat{J}=\sqrt{2J+1}$.  In this expression for the transition matrix elements, the final state ($BC$ system) is written in the ``$J_{bc}$'' instead of the ``$J_c^\prime$'' basis which we use,
\beq
\left[ \left[ J_b J_c \right]_{J_{bc}} l \right]_{J_a}.
\eeq
The initial state is assumed to be in a different basis than the one we use in this work.
\beq
\left[ \left[ J_d s_q \right]_{j_a} l_a \right]_{J_a},
\eeq
where $l_a$ is the relative orbital angular momentum between the heavy degrees of freedom and the light degrees of freedom.  For the case of a DHB, $l_a=l_{\lambda a}$.  Similarly, the daughter heavy hadron is written in a basis,
\beq
\left[ \left[ J_d s_q \right]_{j_b} l_b \right]_{J_b}.
\eeq
Thus, writing the expression for the amplitude in terms of symmetry multiplets requires three separate transformations, each of which consists of a sum over a 6--j symbol:
\begin{eqnarray}
\nonumber {\rm I}   &:&\sum_{J_{bc}} (-1)^{J_b+J_c+l+J_a} \hat{J}_{bc} \hat{J}_c^\prime 
                       \left\{ \begin{array} {ccc} J_b & J_c  & J_{bc} \\ 
                            l & J_a  & J_c^\prime \end{array} \right\} \\
\nonumber {\rm II}  &:& \sum_{j_{a}} (-1)^{J_d+s_q+l_a+J_a} \hat{j}_{a} \hat{J}_{l} 
                       \left\{ \begin{array} {ccc} J_d & s_q  & j_{a} \\ 
                            l_a & J_a  & J_{l} \end{array} \right\} \\
{\rm III} &:& \sum_{j_{b}} (-1)^{J_d+s_q+l_b+J_b} \hat{j}_{b} \hat{J}^\prime_{l} 
                       \left\{ \begin{array} {ccc} J_d & s_q  & j_{b} \\ 
                            l_b & J_b  & J^\prime_{l} \end{array} \right\}  \
\end{eqnarray}

Applying transformation I introduces a sum over $J_{bc}$.  The relevant factor involving this sum is 
\beq \label{JbcSum}      
 \sum_{J_{bc}}  \hat{J}^2_{bc} \left\{ \begin{array} {ccc} j_b & l_b   & J_b \\ 
                            S_c & l_c & J_c \\
                            j_{bc} & l_{bc}  & J_{bc} \end{array} \right\} 
                            \left\{ \begin{array} {ccc} j_{bc} & l_{bc}   & J_{bc} \\ 
                            l & J_a  & L \end{array} \right\}  \\
                           \left\{ \begin{array} {ccc} J_{b} & J_{c}   & J_{bc} \\ 
                            l & J_a  & J_c^\prime \end{array} \right\}.    
\eeq
This sum can be rewritten using a standard identity (see Varshalovich et. al. \cite{Varshalovich}) so as to remove the total angular momentum of the heavy daughter hadron from the 9--j symbol.  Eq.~(\ref{JbcSum}) becomes
\begin{eqnarray}       
\nonumber \sum_X  &\hat{X}^2& \left\{ \begin{array} {ccc}  l_c   & l_{bc} &  l_b \\ 
                                                 J_c   &    l   & J_c^\prime \\
                                                 S_c   &    L   &   X \end{array} \right\} 
                     \left\{ \begin{array} {ccc} l_b & J_c^\prime  & X \\ 
                                                 J_a & j_b  & J_b \end{array} \right\}  
                     \left\{ \begin{array} {ccc} S_{c} & L   & X \\ 
                                                 J_a   & j_b  & j_{bc} \end{array} \right\} \\
         &\times&   (-1)^{j_{bc}+l_{bc}+S_c+L-l_b-J_b-J_c-J_c^\prime}.    
\end{eqnarray}
Applying the same identity to the relevant terms involving the sum over $j_{bc}$ produces
\begin{eqnarray}      
\nonumber \sum_{j_{bc}} &\hat{j}^2_{bc}& \left\{ \begin{array} {ccc} J_d   & \hlf   &  j_b \\ 
                                                           s_q   & \hlf   & S_{c} \\
                                                           j_a   &    1   & j_{bc} \end{array} \right\} 
                     \left\{ \begin{array} {ccc} j_b & S_{c}  & j_{bc} \\ 
                                                 L & J_a  & X \end{array} \right\}  
                     \left\{ \begin{array} {ccc} j_{a} & 1   & j_{bc} \\ 
                                                 L   & J_a  & l_{a} \end{array} \right\} \\
\nonumber = \sum_Y  &\hat{Y}^2& \left\{ \begin{array} {ccc}  \hlf   & \hlf &  1 \\ 
                                                 S_c   &    L   & X \\
                                                 s_q   &   l_a   &   Y \end{array} \right\} 
                     \left\{ \begin{array} {ccc} \hlf & X  & Y \\ 
                                                 J_a & J_d  & j_b \end{array} \right\}  
                     \left\{ \begin{array} {ccc} s_q & l_a   & Y \\ 
                                                 J_a   & J_d  & j_a \end{array} \right\} \\
         &\times&   (-1)^{\thlf+X+j_a-j_b-S_c-s_q-l_a}. \
\end{eqnarray}

If one applies transformation II to the resultant expression for the amplitude, the sum over $j_a$ may be carried out analytically.  The relevant factors involving $j_a$ are,
\begin{eqnarray} \label{jasum}      
\nonumber &&\sum_{j_a} (-1)^{2j_a}  \hat{j}^2_a 
                     \left\{ \begin{array} {ccc} J_d & s_q  & j_a \\ 
                                                 l_a & J_a  & J_l \end{array} \right\}  
                     \left\{ \begin{array} {ccc} J_d & s_q  & j_a \\ 
                                                 l_a   & J_a  & Y \end{array} \right\} \\
         &=&   \frac{\delta_{Y J_l}}{\hat{J}^2_l} \{J_d J_a J_l\} \{s_q l_a J_l\},    
\end{eqnarray}
where 
\beq
\{ abc \}= \left\{ \begin{array}{ll} 1 & \mbox{if } a+b+c \mbox{ is an integer and } |a-b| \le c \le a+b, \\
                         0 & \mbox{otherwise.} 
           \end{array}
           \right.
\eeq
The equality in Eq.~(\ref{jasum}) is exact if $j_a$ is an integer; however, if $j_a$ is a half integer the expression should be modified by an irrelevant overall phase.

Applying transformation III to the remaining expression produces a sum over $j_b$ which may be performed analytically,
\begin{eqnarray} \label{jbsum}      
\nonumber &&\sum_{j_b}  (-1)^{-j_b} \hat{j}^2_b 
                     \left\{ \begin{array} {ccc} J_d & s_q  & j_b \\ 
                                                 l_b & J_b  & J^\prime_l \end{array} \right\}  
                     \left\{ \begin{array} {ccc} l_b & J_b  & j_b \\ 
                                                 J_a   & X  & J_c^\prime \end{array} \right\}
                     \left\{ \begin{array} {ccc} \hlf & J_d  & j_b \\ 
                                                 J_a   & X  & J_{l} \end{array} \right\} \\
  &=&  (-1)^{-(J_d+s_q+l_b+J_b+J_a+X+J_l+J^\prime_l+J_c^\prime)} \left\{ \begin{array} {ccc} J^\prime_l & J_c^\prime  & J_l \\ 
                                                                                             J_a   & J_d  & J_b \end{array} \right\} 
           \left\{ \begin{array} {ccc} J^\prime_l & J_c^\prime & J_l \\ 
                                                 X  & \hlf  & l_b \end{array} \right\}.
\end{eqnarray}
The final expression for the amplitude is proportional to the appropriate spin-counting factor up to an overall phase:
\beq
{\cal M}= (-1)^{J_a+J_d+\hlf} \hat{J}_b \hat{J}_l \left\{ \begin{array} {ccc}  J_d & J^\prime_{l}  & J_b \\ 
                                            J_c^\prime & J_a  & J_l \end{array} \right\} A^{J_c^\prime}_{J_l,J^\prime_l},
\eeq
where
\begin{eqnarray}
\nonumber A^{J_c^\prime}_{J_l,J^\prime_l}&=& 
                     \frac{1}{3} (-1)^{l_a+l_b-J_{l}-J^\prime_{l}} {\cal F} {\cal R} 
                                                                    \frac{\hat{J}^\prime_l \hat{J}_c \hat{S}_c}{\hat{J}_l}
                     \sum_{X} \hat{X}^2 \left\{ \begin{array} {ccc} J_c^\prime & l_b   & X \\ 
                                                            \hlf & J_l  & J^\prime_l \end{array} \right\} \ \\
        &\times& 
 \sum_{l_{bc} L} (-1)^{L} \hat{L}^2 \left\{ \begin{array} {ccc} l_c & l_{bc}   & l_b \\ 
                                                                J_c & l & J_c^\prime \\
                                                                S_c & L  & X \end{array} \right\}
                           \left\{ \begin{array} {ccc} s_q & l_a   & J_l \\ 
                                                       \hlf & \hlf & 1 \\
                                                       S_c & L  & X \end{array} \right\}
                           \epsilon(l_b,l_c,l_{bc},l,l_a,L).     \ \
\end{eqnarray}
By virtue of the generality of the expression for the $^3P_0$ amplitude, Eq.~(\ref{3P0amp}), the derivation above is also valid for heavy mesons if one replaces $J_d$ with $S_Q=\hlf$.  This explicit calculation illustrates the more general spherical tensor argument given in section~\ref{selectrule}.

\input{DHBdecay_bib_mod.tex}

\end{document}

%% file: DHBdecay_bib_mod.tex
\newif\ifmultiplepapers
\def\beginpapers{\multiplepaperstrue}
\def\endpapers{\multiplepapersfalse}  
\def\journal#1&#2(#3)#4{\rm #1~{\bf #2}\unskip, \rm  #4 (19#3)}
\def\trjrnl#1&#2(#3)#4{\rm #1~{\bf #2}\unskip, \rm #4 (19#3)}
\def\baps{\journal {Bull.} {Am.} {Phys.} {Soc.}&}
\def\jap{\journal J. {Appl.} {Phys.}&}
\def\prl{\journal {Phys.} {Rev.} {Lett.}&}
\def\pl{\journal {Phys.} {Lett.}&}
\def\pr{\journal {Phys.} {Rev.}&}
\def\np{\journal {Nucl.} {Phys.}&}
\def\rmp{\journal {Rev.} {Mod.} {Phys.}&}
\def\jmp{\journal J. {Math.} {Phys.}&}
\def\rmm{\journal {Revs.} {Mod.} {Math.}&}
\def\jetp{\journal {J.} {Exp.} {Theor.} {Phys.}&}
\def\sjetp{\trjrnl {Sov.} {Phys.} {JETP}&}
\def\dokl{\journal {Dokl.} {Akad.} Nauk USSR&}
\def\spd{\trjrnl {Sov.} {Phys.} {Dokl.}&}
\def\tmf{\journal {Theor.} {Mat.} {Fiz.}&}
\def\snp{\trjrnl {Sov.} J. {Nucl.} {Phys.}&}
\def\hpa{\journal {Helv.} {Phys.} Acta&}
\def\yf{\journal {Yad.} {Fiz.}&}
\def\zp{\journal Z. {Phys.}&}
\def\anp{\journal {Adv.} {Nucl.} {Phys.}&}
\def\ap{\journal {Ann.} {Phys.}&}
\def\am{\journal {Ann.} {Math.}&}
\def\nc{\journal {Nuo.} {Cim.}&}
\def\etal{{\sl et al.}}
\def\pre{\journal {Phys.} {Rep.}&}
\def\pca{\journal Physica (Utrecht)&}
\def\prs{\journal {Proc.} R. {Soc.} London &}
\def\jcp{\journal J. {Comp.} {Phys.}&}
\def\pna{\journal {Proc.} {Nat.} {Acad.}&}
\def\jpg{\journal J. {Phys.} G (Nuclear Physics)&}
\def\fort{\journal {Fortsch.} {Phys.}&}
\def\jfa{\journal {J.} {Func.} {Anal.}&}
\def\cmp{\journal {Comm.} {Math.} {Phys.}&}